\documentclass[11pt,fleqn]{revtex4}
\usepackage[T1]{fontenc}
\usepackage[latin1]{inputenc}
\usepackage{fancyhdr}
\usepackage[english]{babel}
\usepackage{graphicx}
\usepackage{setspace}
\usepackage{subfigure}
\usepackage{latexsym}
\makeatletter

\begin{document}

\title{Properties and evolution of a protoneutron star in the enlarged SU(3)
model. }

\author{\textbf{Ilona Bednarek, Marcin Keska, Ryszard Manka}}
\address{\textsl{Department of Astrophysics and Cosmology., Institute of Physics,}\\
 \textsl{University of Silesia, Uniwersytecka 4, 40-007 Katowice,
Poland}\\
}

\begin{abstract}
Protoneutron stars are hot and lepton rich objects formed as a result
of type II supernovae explosion. This paper describes results of analysis
of protoneutron star models constructed under the assumptions that
strange particles are present in the core. All calculations have been
performed for the neutrino opaque matter when the entropy per baryon
is of the order 2. The equation of state which is crucial for the
structure and composition of a star has been obtained in the framework
of the chiral SU(3) model and compared with the one involving derivative
coupling of baryons to mesons. Of special interest in this paper is
the comparison of the two protoneutron star models with those formed
when neutrinos leak out of the system, described as cool, deleptonized
neutron stars. It has been shown that the maximum mass of the hot
protoneutron star with trapped neutrinos is close to $1.9M_{\odot}$
in case of the SU(3) model. However, deleptonization reduces the value
of stable protoneutron star mass significantly to $\sim1.5M_{\odot}$.
\end{abstract}
\maketitle

section{Introduction}
The evolution of  a nascent neutron star has to be described under
the assumption of neutrino trapped matter. Numerical solutions
with the lepton number $Y_L \sim 0.4$ are accepted. A protoneutron
star is formed \cite{pra1,pra2,pra3,bur} as a result of a
supernova explosion which leaves  a hot, neutrino opaque core
surrounded by a colder, neutrino transparent outer envelope.
\newline
A protoneutron star structure and  composition depend strongly on
the chosen form of the equation of state which in turn  is
connected with the nature of strong interactions. However, the
character of strong interactions at high density is  still not
 understood completely. At the core of  protoneutron and neutron
stars the matter density ranges from a few times of the density of
normal nuclear matter to the value of one  order higher than that
when hyperons are expected to appear. The direct consequence of the
extreme conditions inside protoneutron and neutron stars is the
possibility of the appearance of different exotic forms of matter.
Of special interest is the existence in these high density
interiors  the strangeness  components like hyperons which may
significantly change the characteristic mass-radius relation of
the star \cite{weber}.  The appearance of the additional degrees
of freedom and their impact on protoneutron star structure and
evolution  have been the subject of extensive studies \cite{glen}.
\newline
The properties of matter at  extreme densities are of particular
importance in determining forms of equations of state relevant to
neutron stars and successively examining their global parameters
\cite{bedn}. Theoretical description of hadronic systems should be
performed with the use of quantum chromodynamics (QCD) as it is
the fundamental theory of strong interactions. However, at the
hadronic energy scale where the experimentally observed degrees of
freedom are not quarks but hadrons the direct description of
nuclei in terms of QCD becomes inadequate. In this paper an
effective model based on chiral symmetry in its nonlinear
realization has been introduced. The chiral SU(3)  model
\cite{sua} which has been applied includes nonlinear scalar and
vector interaction terms. This model offers the possibility of
constructing  a strangeness rich neutron star model and provides
its detail description.
\newline
Another alternative theoretical approach to the correct description
of nuclear matter which has been formulated is quantum
hadrodynamics (QHD) \cite{ser}. This theory  gives quantitative
description of the nuclear many body problem. QHD is a
relativistic quantum field theory in which nuclear matter
description in terms of baryons and mesons is provided. The
original model (QHD-I) contains nucleons interacting through the
exchange of simulating medium range attraction $\sigma$ meson and
$\omega$ meson responsible for short range repulsion. The extension
(QHD-II) of this theory \cite{bog77},\cite{bodmer} includes also
the isovector meson $\rho $. Nonlinear terms in the scalar and
vector fields were added in order to get the correct value of the
compressibility of nuclear matter and the proper density
dependence in the vector self-energy. The variation of nucleon
properties in nuclear medium is the key problem in nuclear
physics. The mentioned above self-consistent relativistic field
models involving coupling of baryons to scalar and vector mesons
are succesfull in describing many properties of nuclear matter.
However, there is a problem connected with the fact that the
nucleon  effective mass becomes very small at moderately high
density \cite{zima}. The situation is even worse when hyperons are
included.
\newline
The onset of hyperon formation depends on the hyperon-nucleon and
hyperon-hyperon  interactions. Hyperons can be formed both in
leptonic and baryonic processes. Several relevant strong
interaction processes proceed and establish the hadron population
in neutron star matter.  When strange hadrons are taken into
account uncertainties which are present in the description of
nuclear matter are intensified due to the incompleteness of the
available experimental data. The standard approach does not
reproduce the strongly attractive hyperon-hyperon interaction seen
in double $\Lambda$ hypernuclei.  In order to construct a proper
model which would include hyperons the effects of hyperon-hyperon
interactions have to be taken into account. These interactions are
simulated via (hidden) strange meson exchange: scalar meson $f_0
(975)$ ($\sigma^*$ meson) and vector meson $\phi (1020)$ ($\phi $
meson) and influence the form of the equation of state and neutron
stars properties.
\newline
 The solution of the presented models are gained
with the mean field approximation in which meson fields are
replaced by their expectation values. The parameters used are
adjusted in the limiting density range around the saturation density
$\rho_0$ and in this density range they give very good description of
finite nuclei. However, incorporation of this theory to higher
density requires an extrapolation which in turn leads to some
uncertainties and suffers from several shortcomings. The standard
TM1 parameter set for high density range reveals an instability of
neutron star matter which is connected with the appearance of
negative nucleon effective mass due to the presence of hyperons.
The Zimanyi-Moszkowski (ZM1) \cite{zima} model in which the Yukawa
type interaction $g_{\sigma N}\sigma$ is replaced by the
derivative one $(g_{\sigma
N}\sigma/M_N)\bar{\psi}_N\gamma_{\nu}\partial^{\nu}\psi$
exemplifies an alternative version of the Walecka model which
improves the behaviour of the nucleon effective masses. It also
influences the value of the incompressibility $K$ of neutron star
matter. The derivative coupling effectively introduces the density
dependence of the scalar and vector coupling constants.
\newline
The chosen models are very useful for describing properties of
nuclear matter and finite nuclei. Its extrapolation to large
charge asymmetry is of considerable interest in nuclear
astrophysics and particulary in constructing protoneutron and
neutron star models where extreme conditions of isospin are
realized \cite{ser,rei,toki,bedn}. The models considered describe
high isospin asymmetric matter and require extension by  the
inclusion of isovector-scalar meson $a_{0}\,(980)$ (the $\delta $
meson) \cite{delta} and   nonlinear meson interaction terms. This
extension
  affects the protoneutron stars chemical composition changing
the proton fraction which in turn affects the properties of the
star.
\section{Properties of protoneutron star matter.}
The collapse of an iron core of a massive star leads to the
formation of a core residue which is considered as an intermediate
stage before the formation of a cold, compact neutron star. This
intermediate stage which is called a protoneutron star can be
described \cite{pra1,pra2,pra3,bur} as
 a hot, neutrino opaque core, surrounded by a colder neutrino
transparent outer envelope. The evolution of a nascent neutron
star  can be described by a series of separate phases starting
from the moment when the star becomes gravitationally decoupled
from the expanding ejecta. In this paper two evolutionary phases
which can be characterized by the following assumptions:
\begin{itemize}
    \item the low entropy core  $s=1-2$ (in units of the Boltzmann's constant) with trapped
    neutrinos $Y_L=0.4$
    \item the cold, deleptonized core ($Y_L=0, s=0$).
\end{itemize}
 have been considered.
These two distinct stages are separated by the period of
deleptonization. During this epoch the neutrino fraction decreases
from the nonzero initial value ($Y_{\nu}\neq 0$) which is
established by the requirement of the fixed  total lepton number
at $Y_l=0.4$,  to the final one characterized by $Y_{\nu}=0$.
Evolution of a protoneutron star which proceeds by neutrino
emission causes that the star changes itself from a hot, bloated
object to a cold, compact neutron star.
\newline
The interior of this  very early stage of a protoneutron star is
an environment in which  matter with the value of entropy of the
order of 2 with trapped neutrinos produces a pressure to oppose
gravitational collapse. The lepton composition of  matter is
specified by the fixed lepton number $Y_L=0.4$. Conditions that
are indispensable for the unique determination of the equilibrium
composition of a protoneutron star matter  arise from the
requirement of $\beta$ equilibrium, charge neutrality and baryon
and lepton number conservation. The later one is strictly
connected with the assumption that the net neutrino fraction
$Y_{\nu}\neq 0$ and therefore the neutrino chemical potential
$\mu_{\nu} \neq 0$. When the electron chemical potential $\mu_e$
reaches the value equal to the muon mass, muons start to appear.
Equilibrium with respect to the reaction
\begin{equation}
e^{-}+\nu_{e} \leftrightarrow \mu^{-}+\bar{\nu}_{\mu}
\end{equation}
is assured when $\mu_{\mu} = \mu_e -\mu_{\nu_{e}}$ (setting
$\mu_{{\nu}_{\mu}}=0$). The appearance of muons reduces the number
of  electrons and also affects  the value of the proton fraction
in  matter. In  the interior of protoneutron stars the density of
matter can substantially exceed the normal nuclear matter density.
In such a high density regime, it is possible that nucleon Fermi
energies exceed the hyperon masses and thus additional hadronic
states are expected to emerge. The higher the density the greater
number of hadronic species are expected to appear. They can be
formed both in leptonic and baryonic processes. The chemical
equilibrium in stellar matter establishes relation between
chemical potentials of protoneutron star matter components.  In
the case when neutrinos are trapped inside matter the requirement
of charge neutrality and equilibrium under the week processes
\begin{equation}
B_1 \rightarrow B_2+f+\bar{\nu}_f \hspace{0.5cm} B_2+f \rightarrow
B_1 +\nu_f
\end{equation}
leads to the following relations
\begin{eqnarray}
\sum_i(n_{B_{i}^{+}}+n_{f^{+}})=\sum_i(n_{B_{i}^{-}}+n_{f^{-}}])\\
\nonumber
 \mu _{i}=b_i\mu _{n}+q_i(\mu _{f}-\mu_{\nu_f})
\end{eqnarray}
where $b_i$ is the baryon number of particle $i$, $q_i$ is its
charge, $f$ stands for leptons $f=e,\mu$ and $\mu_{\nu_f}=\mu_e$.
The nonzero neutrino chemical potential changes chemical
potentials of the constituents of the system and also influences
the onset points and abundance of all species inside the star. The
increase of electron and muon concentration is a significant
effect of neutrino trapping. The proton fraction $Y_p$ also takes
higher value in order to preserve charge neutrality. In the case
of strangeness  rich matter the appearance of charged hyperons
permits the lower electron and muon content and thus the charge
neutrality tends to be guaranteed with the reduced lepton
contribution. After deleptonization the neutrino chemical
potential reduces to zero and chemical equilibrium inside matter
differs from the one presented  above. The relations between
chemical potentials of different constituents of the system is now
obtained setting $\mu_{\nu}=0$. This stage is followed by an
overall cooling stage during which the entropy in the star
decreases.
\section{The model.}
The most general form of the Lagrangian function $\mathcal{L}$ of
the whole system can be shown as a sum of separate parts which
represent baryon together with baryon-meson interactions, scalar,
vector and lepton terms, respectively
\begin{equation}
\mathcal{L}=\mathcal{L}_{B}+\mathcal{L}_{S}+\mathcal{L}_V+\mathcal{L}_L.
\end{equation}
\subsection{Scalar mesons.}
\begin{figure}
\begin{center}\includegraphics[%
  width=8cm]{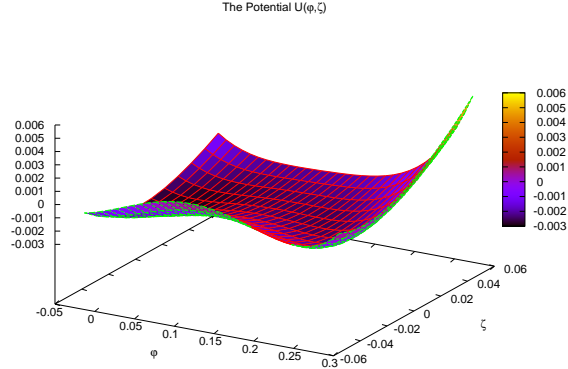}\end{center}
\caption{\it{The form of the potential function
$U(\varphi,\zeta)$.}}
\label{fig:pot1}
\end{figure}

The construction of  the $U(3)_L \times U(3)_R$ invariants
\cite{GG:1969},\cite{NT}  allows us to include different forms of
meson-meson interactions. The following terms have been introduced
\begin{equation}
X=Tr(\Phi^{\dagger}\Phi),
\hspace{0.3cm}Y=Tr[(\Phi^{\dagger}\Phi)^2],
\hspace{0.3cm}Z=det\Phi + det\Phi^{\dagger}.
\end{equation}
where the first two  are $U(3)_L \times U(3)_R$ invariant, whereas
the $Z$ term explicitly breaks the $U(1)_A$ symmetry. The
presented above invariants are constructed from the $3 \times 3$
matrix field $\Phi$ which enables the collective representation of
the  spin $0$ fields
\begin{equation}
\Phi
=\frac{1}{\sqrt{2}}T_a\phi_a=\frac{1}{\sqrt{2}}\lambda_a(\sigma_a+i\pi_a),
\hspace{0.5cm}(a=0,\ldots ,8)
\end{equation}
where $T_a=\lambda_a$ are the generators of $U(3)$ and $\lambda_a$
are the Gell-Mann matrices with
$\lambda_0=\sqrt{\frac{2}{3}}\textbf{1} $ included for the
singlet. The $\lambda$ matrices are normalized by
\begin{equation}
Tr(\lambda_a\lambda_a)=2\delta_{ab}.
\end{equation}
The $\sigma_a$ and $\pi_a$ fields are members of the scalar
$\Sigma=\frac{1}{\sqrt{2}}\lambda_a\sigma_a $ ($J^{P}=0^{+}$)  and
pseudoscalar $\Pi=\frac{1}{\sqrt{2}}\lambda_a\pi_a $
($J^{P}=0^{-}$) nonets respectively. Presenting them as matrices
one can obtain
\begin{equation}
\Sigma=
\left(
\begin{array}{ccc}
 \frac{1}{\sqrt{2}}a_0^0+\frac{1}{\sqrt{6}}\sigma_8+\frac{1}{\sqrt{3}}\sigma_0 & a_0^+ & \kappa^+ \\
 a_0^- & -\frac{1}{\sqrt{2}}a_0^0+\frac{1}{\sqrt{6}}\sigma_8 +\frac{1}{\sqrt{3}}\sigma_0& \kappa^0 \\
  \kappa^- & \bar{\kappa^0} &-\frac{1}{\sqrt{6}}\sigma_8+ \frac{1}{\sqrt{3}}\sigma_0 \\
\end{array}
\right)
\end{equation}
\vspace{0.5cm}
\begin{equation}
\Pi=
\left(
\begin{array}{ccc}
  \frac{1}{\sqrt{2}}\pi^0+\frac{1}{\sqrt{6}}\pi_8+\frac{1}{\sqrt{3}}\pi_0 & \pi^+ & K^+ \\
 \pi^- & -\frac{1}{\sqrt{2}}\pi^0+\frac{1}{\sqrt{6}}\pi_8+\frac{1}{\sqrt{3}}\pi_0  & K^0 \\
   K^- &\overline{K^0} & - \frac{2}{\sqrt{6}}\pi_8+\frac{1}{\sqrt{3}}\pi_0 \\
\end{array}
\right).
\end{equation}
The scalar part of the Lagrangian function $\mathcal{L}_S$ is a
sum of the $U(3)_L \times U(3)_R$ symmetric term and the explicit
symmetry breaking one
\begin{equation}
\mathcal{L}_S(\Phi )= \mathcal{L}_{S,sym} (\Phi
)+\mathcal{L}_{SB}. \label{lag1}
\end{equation}
The $\mathcal{L}_{S,sym}(\Phi )$ part in turn contains the kinetic
and potential terms and  can be expressed as:
\begin{equation}
\mathcal{L}_{S,sym}(\Phi )=
\frac{1}{2}Tr(D_{\mu}\Phi^{\dag}D^{\mu}\Phi) -U(\Phi)
\end{equation}
with the potential function $U(\Phi)$ defined in the following way
\begin{equation}
U(\Phi)= m^2\Phi^{\dag}\Phi +\alpha_1[Tr(\Phi^{\dag}\Phi)]^2
+\alpha_2Tr(\Phi^{\dag}\Phi)^2
\end{equation}
where $m$ is the tree level mass of the fields in the absence of
symmetry breaking term, $\alpha_1$ and $\alpha_2$ are coupling
constants. The explicit symmetry breaking term $\mathcal{L}_{SB}$
has the following form:
\begin{equation}
\mathcal{L}_{SB}=c[Det(\Phi )+Det(\Phi^{\dag})]+Tr[H(\Phi
+\Phi^{\dag})].
\end{equation}
The first term which breaks the $U(1)_A$ symmetry of the
Lagrangian explicitly gives the mass to the pseudoscalar singlet.
In the second term $H$ denotes the $3\times 3$ matrix with
$H=1/\sqrt{2}diag(m_{\pi}^2f_{\pi},m_{\pi}^2f_{\pi},2m_{K}^2f_{K}-m_{\pi}^2f_{\pi})$
where $m_{\pi}=139$ MeV, $m_K=498$ MeV.
\newline
The spontaneous chiral symmetry breaking is triggered by a
non-vanishing expectation value $\overline{\sigma}$ corresponding
to the location of the minimum of the potential $U(\Phi)$. After
the symmetry breaking the $\Phi$ field acquires a vacuum
expectation value. Owing to parity conservation in infinite
neutron star matter the pseudoscalar fields $\pi_a$ cannot assume
a non-vanishing vacuum expectation value for $\Phi$. Shifting the
scalar fields by their vacuum expectation values and substituting
them to the Lagrangian function (\ref{lag1}) allows us to
determine the potential function $U(\Phi-<\Phi>)$.
\newline
The scalar fields $\sigma_a$ in the basis of $U(3)$ generators are
not mass eigenstates and the obtained mass matrix is not diagonal.
Since the mass matrix is real and symmetric, there exists a real
orthogonal matrix $\mathcal{O}$ which diagonalizes $\mathcal{M}$.
On the assumption that $\varphi =(
1/\sqrt{6}\sigma_8+1/\sqrt{3}\sigma_0)$ and $\zeta =
(-1/\sqrt{6}\sigma_8+1/\sqrt{3}\sigma_0)$, the physical scalar
fields are obtained as a result of the following procedure
\begin{equation}
\left\{ \begin{array}{c}
 \varphi \\
\zeta
 \end{array}
\right\}=
\left(%
\begin{array}{cc}
  cos\vartheta & sin\vartheta \\
  -sin\vartheta & cos\vartheta \\
\end{array}%
\right) \left\{ \begin{array}{c}
 \sigma \\
\sigma^{\ast}
 \end{array}
\right\}+
 \left\{ \begin{array}{c}
 \varphi_0 \\
\zeta_0
 \end{array}
\right\}. \label{eq:scalar}
 \end{equation}
 which yields the
diagonalization and shifting the fields to the minimum of the
potential $U(\varphi,\zeta)$.
\newline
The form of the potential function $U(\varphi,\zeta)$ is depicted
in Fig.\ref{fig:pot1}.
For the remaining $a^0_0$ state the following substitution has
been established $a^0_0\equiv \delta $. The field $\sigma$
describes a broad resonance ($m_\sigma\sim470$ MeV) connected with
the exchange of a correlated pair of pions.
\subsection{Baryon-meson interaction.}
Baryon fields that enter the model are grouped into a $3\times 3$
traceless hermitian matrix
\begin{equation}
B=\left(%
\begin{array}{ccc}
  \frac{1}{\sqrt{6}}\Lambda +\frac{1}{\sqrt{2}}\Sigma^{0}& \Sigma^{+} & p \\
  \Sigma^{-} & \frac{1}{\sqrt{6}}\Lambda -\frac{1}{\sqrt{2}}\Sigma^{0} & n \\
  \Xi^{-} & \Xi^{0} & -\frac{2}{\sqrt{6}}\Lambda \\
\end{array}%
\right).
\end{equation}
For the baryon kinetic terms to preserve chiral invariance the
local, covariant derivatives have to be used \cite{sua,zschieshce}
\begin{equation}
D_{\mu}B=\partial_{\mu}+i[\Gamma_{\mu},B]
\end{equation}
where
\begin{equation}
\Gamma_{\mu}=-\frac{i}{2}[u^{\dagger}\partial_{\mu}u+u\partial_{\mu}u^{\dagger}].
\end{equation}
and
\begin{equation}
u=exp[\frac{i}{2f_0}\pi^a\lambda_a\gamma_5].
\end{equation}
Thus the pseudoscalar mesons are given as parameters of the
symmetry transformation.
\newline
The general form of the baryon($B$) meson ($W$) interaction terms
can be written as follows \cite{greiner,sua,zschieshce}
\begin{equation}
\mathcal{L}_{BW}=-\sqrt{2}g_8^W(\alpha_W[\overline{B}\mathcal{O}BW]_F+(1-\alpha_W)[\overline{B}\mathcal{O}BW]_D)-g_1^W\frac{1}{\sqrt{3}}Tr(\overline{B}\mathcal{O}B)TrW
\label{eq:bm}
\end{equation}
where
\begin{equation}
[\overline{B}\mathcal{O}BW]_F=Tr(\overline{B}\mathcal{O}WB-\overline{B}\mathcal{O}BW)\\
\end{equation}
\begin{equation}
[\overline{B}\mathcal{O}BW]_D=Tr(\overline{B}\mathcal{O}WB+\overline{B}\mathcal{O}BW)-\frac{2}{3}Tr(\overline{B}\mathcal{O}B)TrW
\end{equation}
The relation (\ref{eq:bm}) it is a mixture of the F-type
(antisymmetric) and D-type (symmetric) couplings, $\alpha_W$
denotes the $F/F+D$ ratio. The differences for the baryon-scalar
and baryon-vector meson interactions are connected with difference
in Lorentz space. For scalar mesons $W=\Phi$ and
$\mathcal{O}=\textbf{1}$ The masses of the whole baryon multiplet
are generated spontaneously by the vacuum expectation value of the
nonstrange and strange  meson condensates
\begin{equation}
<M>=diag(\frac{\sigma}{\sqrt{2}},\frac{\sigma}{\sqrt{2}},\zeta)
\end{equation}
The assumption that $\alpha_V=1$ has be made. Setting also
$g_1^V=\sqrt{6}g_8^V$  the model in which nucleon mass does not
depend on the strange condensate is obtained. The value of the
nucleon mass can be obtained by adjusting only one coupling
constants. The correct values of the remaining baryons it is
necessary to introduce an explicit symmetry breaking term.
\newline
For the baryon-vector meson interactions $W=V_{\mu}$ and
$\mathcal{O}=\gamma_{\mu}$
 The universality principle and the
vector meson dominance model points to the conclusion that the
D-type coupling should be small. Setting as for the case of scalar
mesons $\alpha_V=1$ and $g_1^V=\sqrt{6}g_8^V$, the latter
assumption corresponds to the case when strange vector field does
not couple to nucleon, the following form of the baryon-vector
meson interaction Lagrangian can be written
\begin{equation}
\mathcal{L}_{BV}=-\sqrt{2}g_8^V([\overline{B}\gamma_{\mu}BV^{\mu}]_F+Tr(\overline{B}\gamma_{\mu}B)TrS^{\mu}).
\end{equation}
with the singlet vector meson state $S_{\mu}$ and $V_{\mu}$ being
the matrix containing the octet of vector meson fields
\begin{equation}V_{\mu}=
\left(%
\begin{array}{ccc}
  \frac{1}{\sqrt{6}}V_{\mu}^8+\frac{1}{\sqrt{2}}\rho^0_{\mu} & \rho_{\mu}^+ & K_{\mu}^{*+} \\
  \rho_{\mu}^- & \frac{1}{\sqrt{6}}V_{\mu}^8-\frac{1}{\sqrt{2}}\rho^0_{\mu} & K_{\mu}^{*0} \\
  K_{\mu}^{*-} & \bar{K}_{\mu}^{*0} & -\frac{2}{\sqrt{6}}V^8_{\mu} \\
\end{array}%
\right).
\end{equation}
The relations for the vector meson couplings reflect quark
counting rules.
\newline
The physical meson states $\omega$ and $\phi$ are mixed states.
They stem from the singlet $S_{\mu}$ and octet $V_{\mu}^8$ vector
meson states and are given by the relation
\begin{eqnarray}
\omega &=& V_{\mu}^8\cos\theta -S_{\mu}\sin\theta \\ \nonumber
\phi &=& V_{\mu}^8\sin\theta +S_{\mu}\cos\theta.
\end{eqnarray}
Under the condition that the vector meson $\phi$ is nearly a pure
$s\overline{s}$ state as it decays mainly to kaons  the mixing
with the mixing angle $\tan{\theta}=1/\sqrt{2}$ is called ideal.
\subsection{Vector mesons.}
The general  vector meson Lagrange function can be presented as a
sum of the kinetic energy term, mass term and higher order
self-interaction terms:
\begin{equation}
\mathcal{L}_V=-\frac{1}{4}Tr(V_{\mu\nu}V^{\mu\nu})+\frac{1}{2}m_V^2Tr(V_{\mu}V^{\mu})-c[Tr(V_{\mu}V^{\mu})^2]-dTr[(V^{\mu}V_{\mu})^2].
\end{equation}
with the field tensor
$V_{\mu\nu}=\partial_{\mu}V_{\nu}-\partial_{\nu}V_{\mu}$. This
form of the mass term implies a mass degeneracy for the meson
nonet.
\subsection{Leptons.}
Leptons  are gathered into $SU_L(2)$ dublets and singlets with the
flavor $f$ ($f={e,\mu}$)
\[
L_{f}=\left(\begin{array}{c}
\nu_{f}\\
e_{f}\end{array}\right)_{L},\,\,\, e_{fR},\,\,\,\nu_{fR}
\]
Assuming that neutrinos $\nu_{f}$ have masses $m_{f}$  the lepton
Lagrange function has the form
\begin{eqnarray*}
 & L_{L}=i\overline{L}_{f}\gamma^{\mu}D_{\mu}L_{f}+\overline{e}_{fR}\gamma^{\mu}D_{\mu}e_{fR}-\sum_{f}m_{f}\overline{e}_{f}e_{f}+\sum_{f}i\overline{\nu}_{fL}\gamma^{\mu}D_{\mu}\nu_{fL}+\sum_{f}i\overline{\nu}_{fR}\gamma^{\mu}\partial_{\mu}\nu_{fR}\\
 & -\sum_{f,f'}M_{f,f'}(\overline{\nu}_{fL}\nu_{f'R}+h.c.)-\sum_{f,f'}M_{f,f'}(\overline{\nu}_{fL}\nu_{f'R}+h.c.)
 \end{eqnarray*}
 The physical massive neutrinos are mixture of neutrinos with
different flavors
\begin{eqnarray*}
 & \nu_{fL}=\sum_{i}U_{L,fi}\nu_{iL}\\
 & \nu_{fR}=\sum_{i}U_{R,fi}\nu_{iR}
\end{eqnarray*}
\section{Relativistic mean field equations.}
\begin{figure}
\begin{center}
\includegraphics[width=8cm]{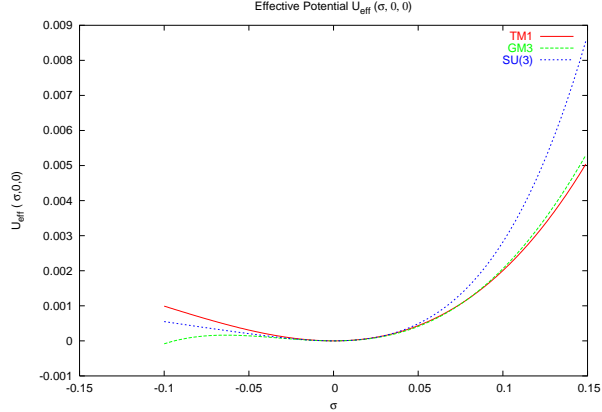} \caption{\it{The scalar
potential function calculated for different parameter sets.}}
\label{fig:potsc}
\end{center}
\end{figure}

\begin{figure}
\begin{center}
\includegraphics[width=8cm]{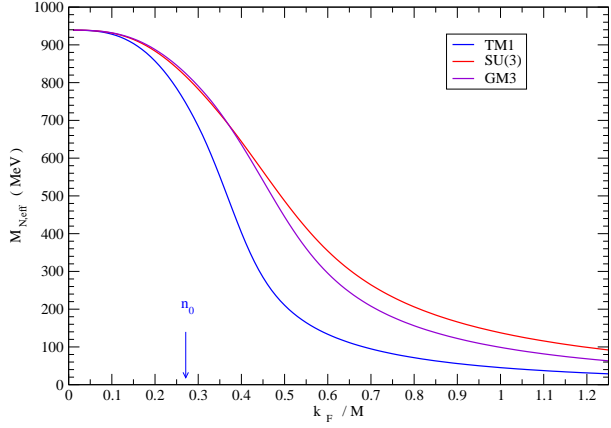} \caption{\it{The effective
nucleon masses  obtained for different parameter sets.}}
\label{fig:meff}
\end{center}
\end{figure}

To investigate the properties of infinite nuclear matter, the mean
field approximation has been adopted.  The symmetries of infinite
nuclear matter simplify the model to a great extent. The
translational and rotational invariance claimed that the mean
fields of all the vector fields vanish.  Only the time-like
components of the neutral vector mesons have a non-vanishing
expectation value. Owing to parity conservation, the vacuum
expectation value of pseudoscalar fields vanish ($<\pi_a>=0$).
Meson fields have been separated into classical mean field values
and quantum fluctuations, which are not included in the ground
state. Thus, for the ground state of homogeneous infinite nuclear
matter quantum fields operators are replaced by their classical
expectation values.
\begin{center}
\begin{tabular}{lll}\\
  \hline
$\sigma  = \overline{\sigma}$ +$ s_0$ &$ \sigma^*  =
\overline{\sigma}^*$
+ $s^*_0$ & $\delta^a = \overline{\delta}^a+ d_0\delta^{3a}$  \\
$\phi_{\mu}  = \overline{\phi}_{\mu} + f_0\delta_{\mu 0}$ &
$\omega_{\mu} = \overline{\omega}_{\mu}+ w_{0}\delta_{\mu 0}$ &
$b_{\mu}^a  =  \overline{b}^a_{\mu}+r_{0}\delta_{\mu
0}\delta^{3a}$ \\ \hline
\end{tabular}
\end{center}
When strange hadrons are taken into account uncertainties which
are present in the description of nuclear matter are intensified
due to the incompleteness of the available experimental data. The
standard approach does not reproduce the strongly attractive
hyperon-hyperon interaction seen in double $\Lambda$ hypernuclei.
In order to construct a proper model which do include hyperons the
effects of hyperon-hyperon interactions have to be taken into
account. These interactions are simulated via (hidden) strange
meson exchange: scalar meson $f_0 (975)$ ($\sigma^*$ meson) and
vector meson $\phi (1020)$ ($\phi $ meson) and influence the form
of the equation of state and neutron stars properties. The
presence of hyperons demands additional coupling constants which
have been fitted to hypernuclear properties.
\newline
 The Lagrangian function for the system is a sum of a baryonic
part, including the full octet of baryons, baryon-meson
interaction terms, a mesonic part and a free leptonic one. The
Lagrangian density of interacting byarons ${\mathcal{L}_{BM}}$ can
be explicitly written out in the following form:
\begin{equation}
{\mathcal{L}_{BM}}=i\sum_B\bar{\psi}_B\gamma^{\mu}D_{\mu}\psi_B-\sum_B\overline{\psi}_BM_{eff,B}\psi_B
\end{equation}
where  the spinor $\Psi^T
=(\psi_N,\psi_{\Lambda},\psi_{\Sigma},\psi_{\Xi})$ is composed of
the following isomultiplets \cite{glen},\cite{aqu}:
\[\begin{array}{cc}
\displaystyle\Psi_N={\psi_p \choose \psi_n}, &
\displaystyle\Psi_{\Lambda}=\psi_{\Lambda},\\ \\
\displaystyle\Psi_{\Sigma}=\left(
\begin{array}{c}\psi_{\Sigma^+}\\ \psi_{\Sigma^0} \\
\psi_{\Sigma^-} \end{array}\right),&
\displaystyle\Psi_{\Xi}={\psi_{\Xi^0} \choose \psi_{\Xi^-}}.
\end{array}\]
$D_{\mu}$ is the covariant derivative  given by
\begin{equation}
D_{\mu}=\partial_{\mu}+ig_{\omega B}\omega_{\mu}+ig_{\phi
B}\phi_{\mu}+ig_{\rho B}I_{3B}\tau^a\rho^a_{\mu}.
\end{equation}
The  meson part of the Lagrangian function includes nonlinear
terms that describe additional interactions between mesons.
\begin{eqnarray}
\label{lagrangian}
  \mathcal{L}_{M}=
 \frac{1}{2}\partial_{\mu}\sigma\partial^{\mu}\sigma+\frac{1}{2}\partial_{\mu}\delta^a\partial^{\mu}\delta^a+
 \frac{1}{2}\partial_{\mu}\sigma^{\ast}\partial^{\mu}\sigma^{\ast}- U_{eff,s}(\sigma,\delta,\sigma^{\ast})
 & & \nonumber \\ -\frac{1}{4}\Omega_{\mu \nu}\Omega^{\mu \nu}+\frac{1}{2}
M_{\omega }^{2}\omega_{\mu}\omega^{\mu}
-\frac{1}{4}R^a_{\mu\nu}R^{a\mu\nu}+\frac{1}{2}M_{\rho
}^{2}\rho^a_{\mu}\rho^{a\mu}& & \nonumber \\
-\frac{1}{4}\Phi_{\mu\nu}\Phi^{\mu\nu}+\frac{1}{2}M_{\phi}^2\phi_{\mu}\phi^{\mu}
+U_{eff,v}(\omega,\rho,\phi).
\end{eqnarray}
The field tensors $\Omega_{\mu\nu}, \Phi_{\mu\nu}$ and
$R^{a}_{\mu\nu}$ are defined as
\begin{equation}
\Omega_{\mu\nu}=\partial_{\mu}\omega_{\nu}-\partial_{\nu}\omega_{\mu}
\hspace{0.5cm}\Phi_{\mu\nu}=\partial_{\mu}\phi_{\nu}-\partial_{\nu}\phi_{\mu}
\end{equation}
\begin{equation}
R^{a}_{\mu\nu}=\partial_{\mu}\rho^{a}_{\nu}-\partial_{\nu}\rho^{a}_{\mu}+g_{\rho}\epsilon_{abc}\rho^{b}_{\mu}\rho^{c}_{\nu}.
\end{equation}
All nonlinear meson interaction terms are collected in the vector
and scalar potential functions $U_{eff,v}(\omega,\rho,\phi)$ and
$U_{eff,s}(\sigma
,\delta,\sigma^{\ast})$.\\
The form of the scalar potential is portrayed   by
\begin{eqnarray}
U_{eff,s}(\sigma,\delta,
\sigma^{\ast})=\frac{1}{2}m_{\sigma}^2\sigma^2+\frac{1}{3}g_2\sigma^3+\frac{1}{4}g_3\sigma^4+ \frac{1}{2}m_{\delta}^2(\delta^a\delta^a)    \\
\nonumber +\frac{1}{4}g_{\delta
4}(\delta^a\delta^a)^2+g_{\sigma\delta
2}\sigma\delta^a\delta^a+\frac{1}{2}g_{\sigma 2\delta
2}\sigma^2\delta^a\delta^a \\ \nonumber +
\frac{1}{2}m_{\sigma\ast}^2\sigma^{\ast2} +
\frac{1}{3}g_{\sigma\ast3}\sigma^{\ast3}+\frac{1}{4}g_{\sigma\ast4}\sigma^{\ast4}+g_{\sigma
\sigma\ast2}\sigma\sigma^{\ast2} \\  \nonumber +g_{\sigma
\sigma\ast3}\sigma\sigma^{\ast3} +g_{\sigma2
\sigma\ast}\sigma^2\sigma^{\ast}+g_{\sigma2
\sigma\ast2}\sigma^2\sigma^{\ast2}+g_{\sigma3
\sigma\ast}\sigma^3\sigma^{\ast} \\ \nonumber + g_{\delta2
\sigma\ast}\delta^2\sigma^{\ast}+g_{\delta2
\sigma\ast2}\delta^2\sigma^{\ast2}+g_{\sigma\delta2\sigma\ast}\sigma\delta^2\sigma^{\ast},
\end{eqnarray}
whereas the effective vector potential has the following form
\begin{equation}
U_{eff,v}(\omega,\rho,\phi)=\frac{1}{4}c_3\left((\omega_{\mu}\omega^{\mu})^2+(\rho^a_{\mu}\rho^{a\mu})^2
-3(\omega_{\mu}\omega^{\mu}+\rho^{a}_{\mu}\rho^{a\mu})(\phi_{\mu}\phi^{\mu})
 -(\phi_{\mu}\phi^{\mu})^2\right)
\end{equation}
The potential function $U_{eff,s}(\sigma,0,0)$ is presented in
Fig.\ref{fig:potsc}.
Scalar potential functions generated for TM1 and GM3 (Glendenning
and Moszkowski \cite{GM:1991}) parameter sets are also included in
this figure.
\\
\begin{table}
\begin{center} \caption{Parameter sets for chosen equations of
state.}\label{tab:TM1}
\begin{tabular}{|c|c|c|c|} \hline & TM1 &
GM3 & SU3 \tabularnewline \hline \hline $m_{\sigma}$ (MeV) &511.2
& 450.0 & 477.6  \tabularnewline \hline $m_{\delta}$ (MeV) & 980.0
& 980.0 & 1029  \tabularnewline \hline $g_{2}$ & 7.2325 $fm^{-3}$
& 15.286 $fm^{-3}$ & 16.196 $fm^{-3}$  \tabularnewline \hline
$g_{3}$ & 0.6183 & -6.4547 & 15.669  \tabularnewline \hline
$g_{\sigma N}=g_{\sigma}$ & 10.029 & 7.186 & 7.923 \tabularnewline
\hline $g_{\omega}$ & 12.61 & 8.702 & 9.225 \tabularnewline \hline
$g_{\rho}$ & 9.264 & 8.542 & 5.742 \tabularnewline \hline
$g_{\sigma\delta2}$ & 0 & 0 & 32.639 $fm^{-3}$  \tabularnewline
\hline $g_{\sigma2\delta2}$ & 0 & 0 & 29.511  \tabularnewline
\hline $g_{\delta4}$ & 0 & 0 & 19.781 \tabularnewline \hline
\end{tabular}
\end{center}
\end{table}
\begin{table}
\begin{center}\caption{The nonlinear SU3 parameter set.} \label{tab:SU3}
\begin{tabular}{|c|c|c|c|c|} \hline
 $ m_{\sigma}$(MeV) & $m_{\delta}$(MeV) & $g_2$ & $g_3$ & $g_{\sigma}$ \\
  \hline
  477.6 & 1029 & 16.196 $fm^{-3}$ & 15.669 & 7.923 \\\hline \hline
$g_{\omega}$ & $g_{\rho}$ & $g_{\delta4}$ & $g_{\sigma\delta2}$ &
$g_{\sigma2\delta2}$
\\\hline
  9.225 & 6.754 & 19.78 & 31.69 $fm^{-1}$ & 59.02 \\\hline \hline
  $g_{\sigma\ast3}$ & $g_{\delta2\sigma\ast}$ & $g_{\delta2\sigma\ast2}$ & $g_{\sigma\delta2\sigma\ast}$ & $g_{\sigma\ast4}$
  \\\hline
  49.88$fm^{-1}$ & -9.51 & 1.90 & 21.39 & 34.07 \\ \hline \hline
  $g_{\sigma\sigma\ast2}$ & $g_{\sigma\sigma\ast3}$ & $g_{\sigma2\sigma\ast}$ & $g_{\sigma2\sigma\ast2}$ & $g_{\sigma3\sigma\ast}$
  \\\hline
  25.27 & 12.01 & -9.51 & 1.90 & 4.88 \\
  \hline \hline
\end{tabular}
\end{center}
\end{table}
The scalar $\sigma$ coupling constants for hyperons are chosen to
reproduce hyperon potentials in saturated nuclear matter:
\begin{equation}
U_{\Lambda}^{N}(\rho_0)=-30
MeV,\hspace{0.5cm}U_{\Sigma}^{N}(\rho_0)=+30
MeV,\hspace{0.5cm}U_{\Xi}^{N}(\rho_0)=-18 MeV
\end{equation}
Upon a recent analysis of $\Sigma^-$ atomic data there has been
indication of the existence of a repulsive isoscalar potential in
the interior of nuclei. This corresponds to a resent search for
$\Sigma$ hypernuclei which show  the lack of bound-state or
continuum peaks.  The only found $\Sigma$ bound state is
${}^{4}_{\Sigma}$He. The binding results from the strong isovector
component of the $\Sigma$ nuclear interaction. Thus, in the
considered model $\Sigma$ hyperons will not appear in the neutron
star interior. The experimental data concerning hyperon-hyperon
\cite{dhsf} interaction are extremely scarce. The observed double
$\Lambda$ hypernuclear events in emulsion require strong
$\Lambda\Lambda$ attractive interaction. An analysis of events
which can be interpreted as the creation of $\Lambda \Lambda$
hypernuclei allows us to determine the well depths of hyperon in
hyperon matter. The $\sigma^{\ast}$ coupling constants to
hyperons, in accordance with the one-boson exchange model D of the
Nijmegen group and the measured strong $\Lambda\Lambda$
interaction, are fixed by the condition:
\begin{equation}
-U_{\Xi}^{(\Xi)}\approx
-U_{\Lambda}^{(\Xi)}\approx-2U_{\Xi}^{(\Lambda)} \approx
-2U_{\Lambda}^{(\Lambda)}\approx 40 MeV.
\end{equation}
The appropriate parameter set is denoted as SU3 constrained not
only by the value of physical scalar meson masses but also by the
properties of nuclear matter at saturation. The obtained results
are collected in Table 1 and 2.\\
 The effective interaction is
introduced through Klain-Gordon equations for the meson fields
with baryon densities as source terms. These equations are coupled
to the Dirac equations for baryons. All the  equations have to be
solved self-consistently. The field equations derived from the
Lagrange function at the mean field level are the following:
\begin{eqnarray}
(m_\sigma^2+g_{\sigma2\delta2}d_0^2+g_{\sigma2\sigma\ast}s_0^{\ast})s_0+(g_2+g_{\sigma3\sigma\ast}s_0^{\ast})s_0^2+g_3s_0^3
\\ \nonumber =\sum_Bg_{\sigma B}M^2_{eff}S(M_{eff,B})
\end{eqnarray}
\begin{equation}
m_{\omega}^2w_{0}+ c_3w_{0}^3-\frac{3}{2}c_3w_0f_0^2
=\sum_Bg_{\omega B}n_B
\end{equation}
\begin{equation}
m_{\rho}^2r_0+c_3r_0^3-\frac{3}{2}c_3r_0f_0=\sum_Bg_{\rho
B}I_{3B}n_B
\end{equation}
\begin{eqnarray}
m_{\delta}^2d_0^3+2(g_{\sigma\delta2}s_0+\frac{1}{2}g_{\sigma2\delta2}s_0^2+\frac{1}{4}g_{\delta4}+g_{\delta2\sigma\ast}s_0^{\ast}+g_{\delta2\sigma\ast2}s_0^{\ast2}+g_{\sigma\delta2\sigma{\ast}}s_0^{\ast}s_0)d_0
\\ \nonumber =\sum_Bg_{\delta B}I_{3B}S(M_{eff,B})
\end{eqnarray}
\begin{eqnarray}
m_{\sigma*}^2s_0^*+(g_{\sigma\ast3}+3g_{\sigma\sigma\ast3}s_0)s_0^{\ast2}+g_{\sigma\ast4}s_0^{\ast3}+2(g_{\sigma\sigma\ast2}s_0+g_{\sigma2\sigma\ast2}s_0^2+g_{\delta2\sigma\ast2}d_0^2)s_0^{\ast}
\\ \nonumber
+(g_{\sigma2\sigma\ast}s_0^2+g_{\sigma3\sigma\ast}s_0^3+g_{\delta2\sigma\ast}d_0^2)=\sum_Bg_{\sigma^*B}M^2_{eff}S(M_{eff,B})
\end{eqnarray}
\begin{equation}
m_{\phi}^2f_0-\frac{3}{2}c_3(w_0^2+r_0^2)f_0-c_3f_0^3=\sum_Bg_{\phi
B}n_B.
\end{equation}
The function $S(M_{eff,B})$ is expressed with the use of the
integral
\begin{equation}
S(M_{eff,B})=\frac{2J_B+1}{2\pi^2}\int_0^\infty
\frac{k^2dk}{E(k,M_{eff,B})}(f_B-f_{\bar{B}})
\end{equation}
where $J_B$ and $I_{3B}$ are the spin and isospin projection of
baryon $B$, $n_B$ which denote the baryon number density is given
as
\begin{equation}
n_B=\frac{2J_B+1}{2\pi^2}\int_0^\infty k^2dk(f_B-f_{\bar{B}})
\end{equation}
The functions $ f_B$ and $f_{\bar{B}} $ are the Fermi-Dirac
distribution for particles and anti-particles respectively
\begin{equation}
f_{B,\bar{B}}=\frac{1}{1+e^{E[(k,M_{eff,B})\mp \mu_B]/k_BT}}.
\end{equation}
 The Dirac equation for  baryons that is
obtained from the Lagrangian function has the following form:
\begin{equation}
(i\gamma ^{\mu }\partial_{\mu }-M_{B,eff}-g_{\omega
B}\gamma^{0}\omega_{0}-g_{\phi
B}\gamma^{0}f_{0}-\frac{1}{2}g_{\rho B}\gamma^{0}\tau^3r_{0})\psi
=0
\end{equation}
from this equation it is noticeable that the baryon develop the
effective mass $M_{B,eff}$  generated by the baryon and scalar
field interactions and  defined as:
\begin{equation}
M_{B,eff}=M_B-(g_{\sigma
B}s_0+g_{\sigma^*B}s_0^*+I_{3B}g_{\delta_B}d_0)
\label{effectivemass}
\end{equation}
 and the effective chemical potential $\mu_{eff}$ is given by
\begin{equation}
\mu_{B,eff}=\mu_B-(g_{\omega B}\gamma^{0}\omega_{0}-g_{\phi
B}\gamma^{0}f_{0}-\frac{1}{2}g_{\rho B}\gamma^{0}\tau^3r_{0})
\end{equation}
Numerical solutions of the equation (\ref{effectivemass})  is
presented in Fig.\ref{fig:meff}.
The main effect of the inclusion of $\delta$ meson itself becomes
evident when  studying  properties of  neutron star matter
especially baryon mass splitting and the form of the equation of
state. Fig.\ref{fig:meff}  depicts the effective baryon masses as
a function of baryon number density $n_B$. There is a noticeable
mass splitting of baryons  for each isomultiplet in medium which
depends on the considered baryonic masses. The bigger is the mass,
the smaller the mass differences are. Depending on the sign of the
third component of particular baryon isospin the $\delta$ meson
interaction increases the proton and $\Xi^0$ effective masses and
decreases masses of neutron, $\Sigma^-$ and $\Xi^-$. Contrary to
this situation, when $\delta $ meson is not included, the baryon
mass for a given isomultiplet remains degenerated.  Throughout the
effective baryonic masses $\delta$ meson alters baryon chemical
potentials what is reflected in characteristic modification of the
appearance, abundance and
distributions of the individual flavors.\\

\begin{table}
\begin{center}
\caption{Meson-hyperon coupling constants for the ZM1
model.}\label{tab:hipr} {\begin{tabular}{|c|c|c|c|c|c|}
  \hline
$g_{\sigma\Lambda}$ & 0.5207 $g_{\sigma N}$ &
$g_{\sigma^{\ast}\Lambda}$ & 0.5815 $g_{\sigma
N}$&$g_{\omega\Lambda}$ & $\frac{2}{3}$ $g_{\omega N}$ \\\hline
$g_{\rho\Lambda}$  & 0 & $g_{\delta\Lambda}$ &  0 &
$g_{\phi\Lambda}$ & -$\frac{\sqrt{2}}{3}$ $g_{\omega N}$\\\hline
$g_{\sigma\Sigma}$ & 0.1565 $g_{\sigma N}$ &
$g_{\sigma^{\ast}\Sigma}$ & 0 &$g_{\omega\Sigma}$ & $\frac{2}{3}$
$g_{\omega N}$\\ \hline $g_{\rho\Sigma}$ & 2 $g_{\rho N}$&
$g_{\delta\Sigma}$ & 2 $g_{\delta N}$ & $g_{\phi\Sigma}$ &
-$\frac{\sqrt{2}}{3}$ $g_{\omega N}$ \\ \hline $g_{\sigma\Xi}$ &
0.2786 $g_{\sigma N}$ & $g_{\sigma^{\ast}\Xi}$ & 0.5815 $g_{\sigma
N}$ &$g_{\omega\Xi}$ & $\frac{1}{3}$ $g_{\omega N}$\\\hline
$g_{\rho\Xi}$& $g_{\rho N}$ & $g_{\delta\Xi}$ & $g_{\delta N}$&
$g_{\phi\Xi}$ &
-$\frac{2\sqrt{2}}{3}$ $g_{\omega N}$ \\
\hline
\end{tabular}}
\end{center}
\end{table}
As it was stated earlier neutron star matter possesses highly
asymmetric character caused by the presence of small amounts of
protons and electrons. The introduction of the asymmetry parameter
$f_a$  which describes the relative neutron excess defined as
\begin{equation}
f_a=\frac{n_n-n_p}{n_N}
\end{equation}
allows to study the symmetry properties of the system. Similarly
to the asymmetry parameter $f_a$  a parameter $f_s$ which specify
the strangeness content in the system and is strictly connected
with the appearance of particular hyperon species in the model has
been introduced.
\begin{equation}
f_s=\frac{n_{\Lambda}+n_{\Sigma}+2n_{\Xi}}{n_{\Lambda}+n_{\Sigma}+n_{\Xi}+n_{N}}.
\end{equation}
\section{The derivative coupling model.}
 The description of the nuclear matter properties
with the use of the standard TM1 parameter set reveals a
shortcoming which is connected with the appearance of negative
nucleon effective masses for densities characteristic for hyperon
stars. The Zimanyi-Moszkowski (ZM1) \cite{zima} model in which the
Yukawa type interaction $g_{\sigma N}\sigma$ is replaced by the
derivative one $(g_{\sigma
N}\sigma/M_N)\bar{\psi}_N\gamma_{\nu}\partial^{\nu}\psi$
exemplifies an alternative version of the Walecka model and
improves the behavior of the nucleon effective masses.
\newline
In this model in the term which
  represents the  Lagrangian
density of interacting baryons is given as follows
\begin{equation}
{\mathcal{L}_{BM}}=\sum_{B}(1+\frac{g_{\sigma B}\sigma
+g_{\sigma^*B}\sigma^*+I_{3B}g_{\delta
B}\tau^a\delta^a}{M_B})\bar{\psi}_Bi\gamma^{\mu}D_{\mu}\psi_B-\sum_B\overline{\psi}_BM_B\psi_B
\end{equation}
 Rescaling the baryon field in a way proposed by Zimanyi and
Moszkowski \cite{zima} the modified Lagrange function for
interacting baryons is obtained
\begin{equation}
{\mathcal{L}_{B}}=-\sum_B\bar{\psi}_Bi\gamma^{\mu}D_{\mu}\psi
-\sum_B\left(1+\frac{g_{\sigma B}\sigma
+g_{\sigma^*B}\sigma^*+I_{3B}g_{\delta
B}\tau^a\delta^a}{M_B}\right)^{-1}\overline{\psi}_BM_B\psi_B
\end{equation}
Expanding the expression $(1+g_{\sigma N}\sigma/M_N)^{-1}$ in
terms of $ g_{sN}\sigma/M_N$ (index $N$ denotes nucleons) up to
first order in $\sigma$ allows one to reproduce the baryonic part
of the Lagrangian of the Walecka model with a Yukawa $N-\sigma$
interaction. The effective baryon mass in this case is given by
the following formula
\begin{equation}
M_{B,eff}=\frac{M_B}{1+(g_{\sigma
B}\sigma+g_{\sigma^*B}\sigma^*+I_{3B}g_{\delta_B}\tau^a\delta^a)/M_B}
\end{equation}
The parameters employed in the ZM1  model are collected in Tables
3 and 4 \cite{bedn2}. In this case the value of the parameter
$g_{\rho N}$ has to be redefined in comparison with the standard
TM1 value. The parameters $g_{\rho N}$ and $g_{\delta N}$ are
adjusted to obtain the symmetry energy coefficient $a_{sym}(n_0)$
at saturation equal $32.4$ MeV which is in good agreement with the
empirical value of about $30 \pm 4$ MeV.
\begin{table}
\begin{center}\label{tab:par}
\caption{Meson-nucleon coupling constants for the ZM1 model.}
{\begin{tabular}{|c|c|c|c|}
  \hline
  $g_{\sigma N}$ & 7.84 &$g_{\phi N}$ & 0 \\ \hline
  $g_{\omega N}$ & 6.671 & $g_{3}$ & 0 \\ \hline
  $g_{\rho N}$ & 9.5 & $g_{4}$ & 0 \\ \hline
  $g_{\delta N}$ & 3.1 & $c_{3}$ & 0 \\ \hline
  $g_{\sigma^{\ast}N}$ & 0 &  &\\\hline
\end{tabular}}
\end{center}
\end{table}
\section{Results and conclusions.}
\begin{figure}[htbp]
\begin{center}
\includegraphics[width=7cm]{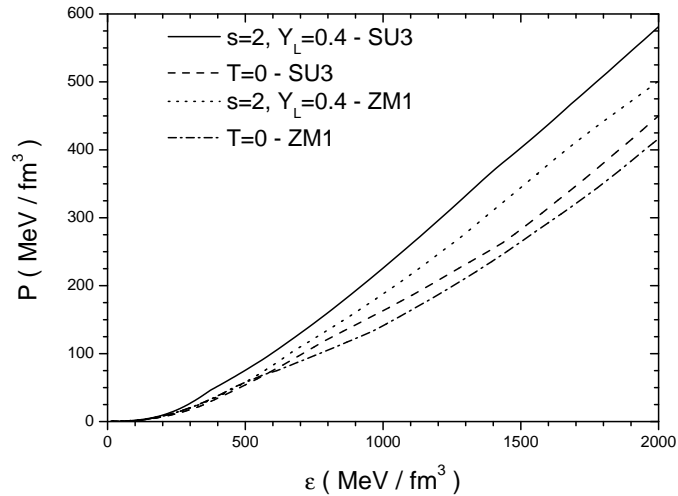}
\end{center}
\caption{{\it The pressure energy density relations for the
selected parameter sets.}}
 \label{fig:eos}
\end{figure}

\begin{figure}
\begin{center}
\subfigure{\includegraphics[width=6cm]{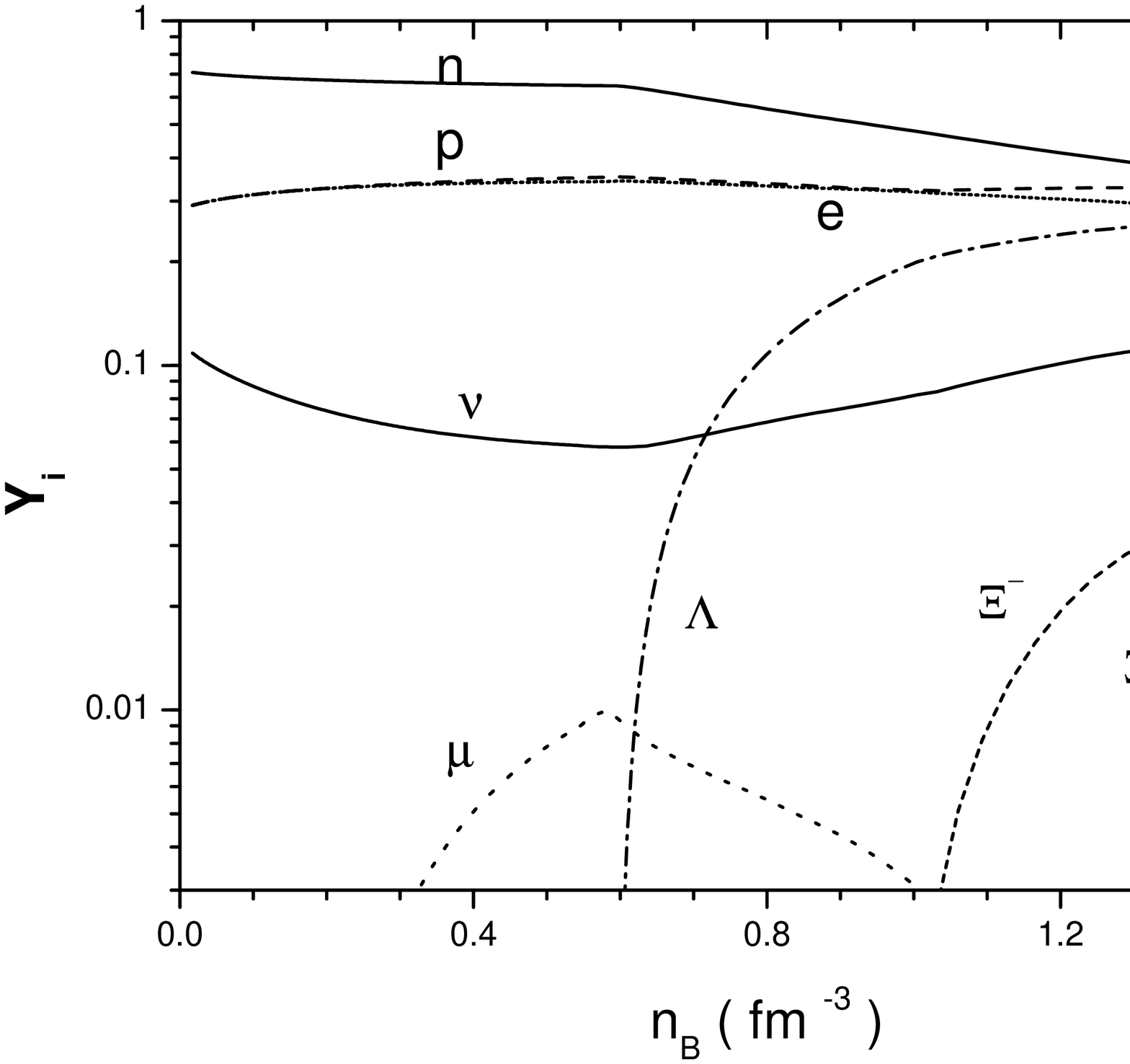}}
~~~~~~~~~~~~
\subfigure{\includegraphics[width=6cm]{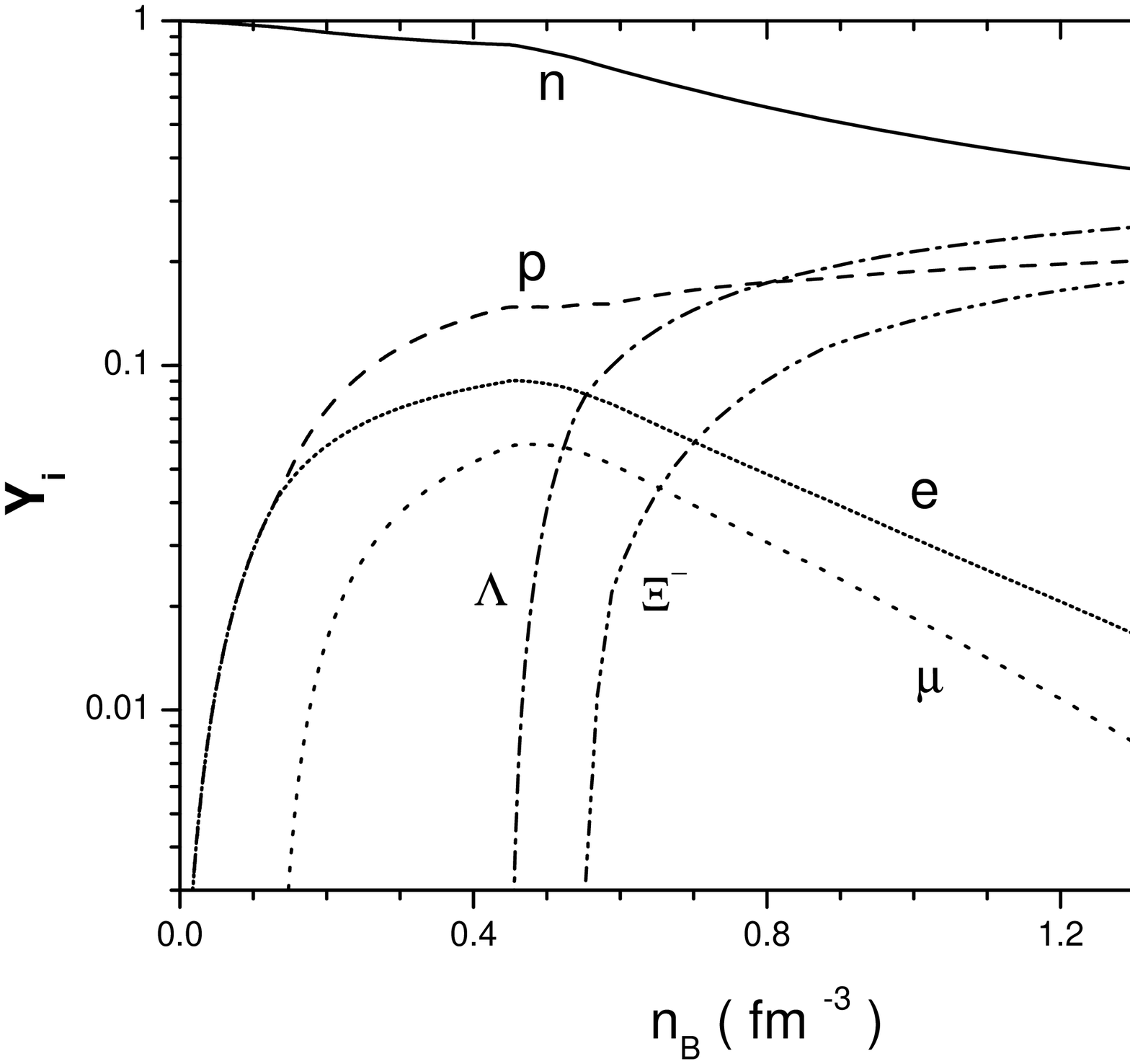}}
\subfigure{\includegraphics[width=6cm]{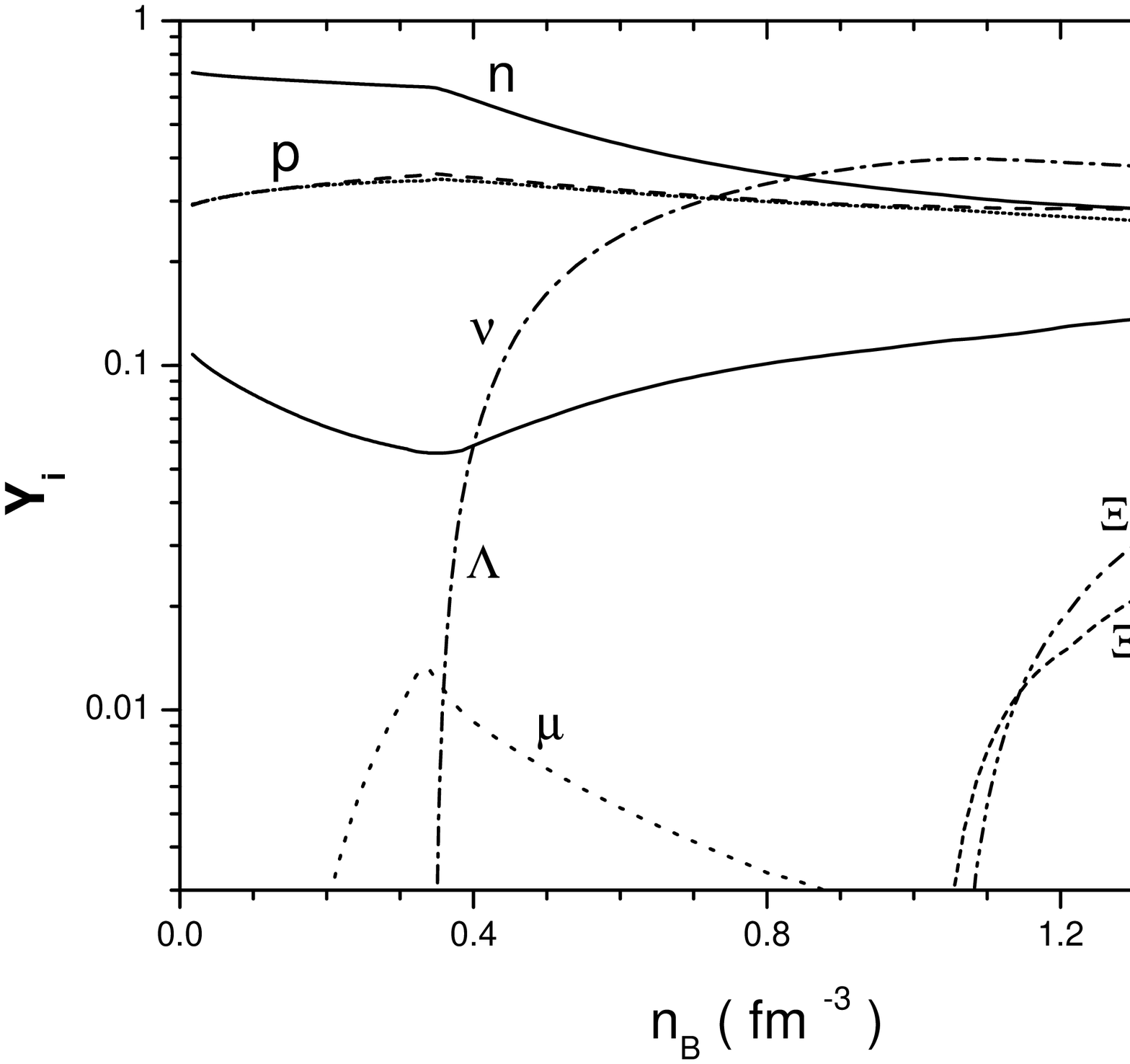}}
~~~~~~~~~~~~
\subfigure{\includegraphics[width=6cm]{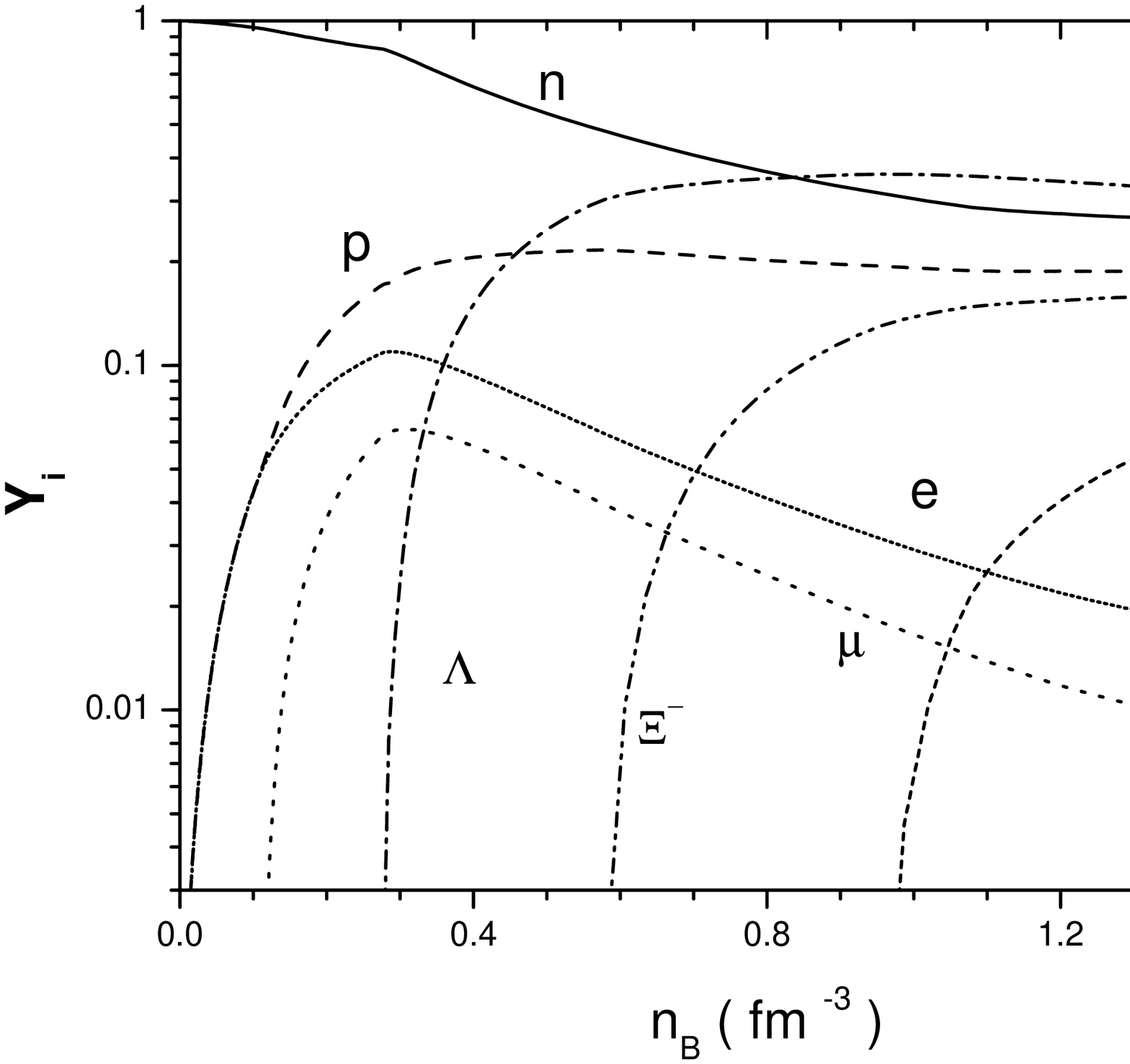}}
\end{center}
\caption{{\it {The equilibrium composition of protoneutron star
matter as  a function of the baryon number density 
$n_{B}$. 
Upper left panel depicts the composition obtained on the basis of the
ZM1 model for $s=2$ and $Y_l=0.4$. Upper right has been calculated
for $s=0$ and $Y_l=0$. Results obtained for SU3 model are
presented in lower panels $s=2$ and $Y_l=0.4$ (left) and $s=0$ and
$Y_l=0$(right).}}}
 \label{fig:partn}
\end{figure}

\begin{figure}[htbp]
 \begin{center}
\subfigure {\includegraphics[width=6cm]{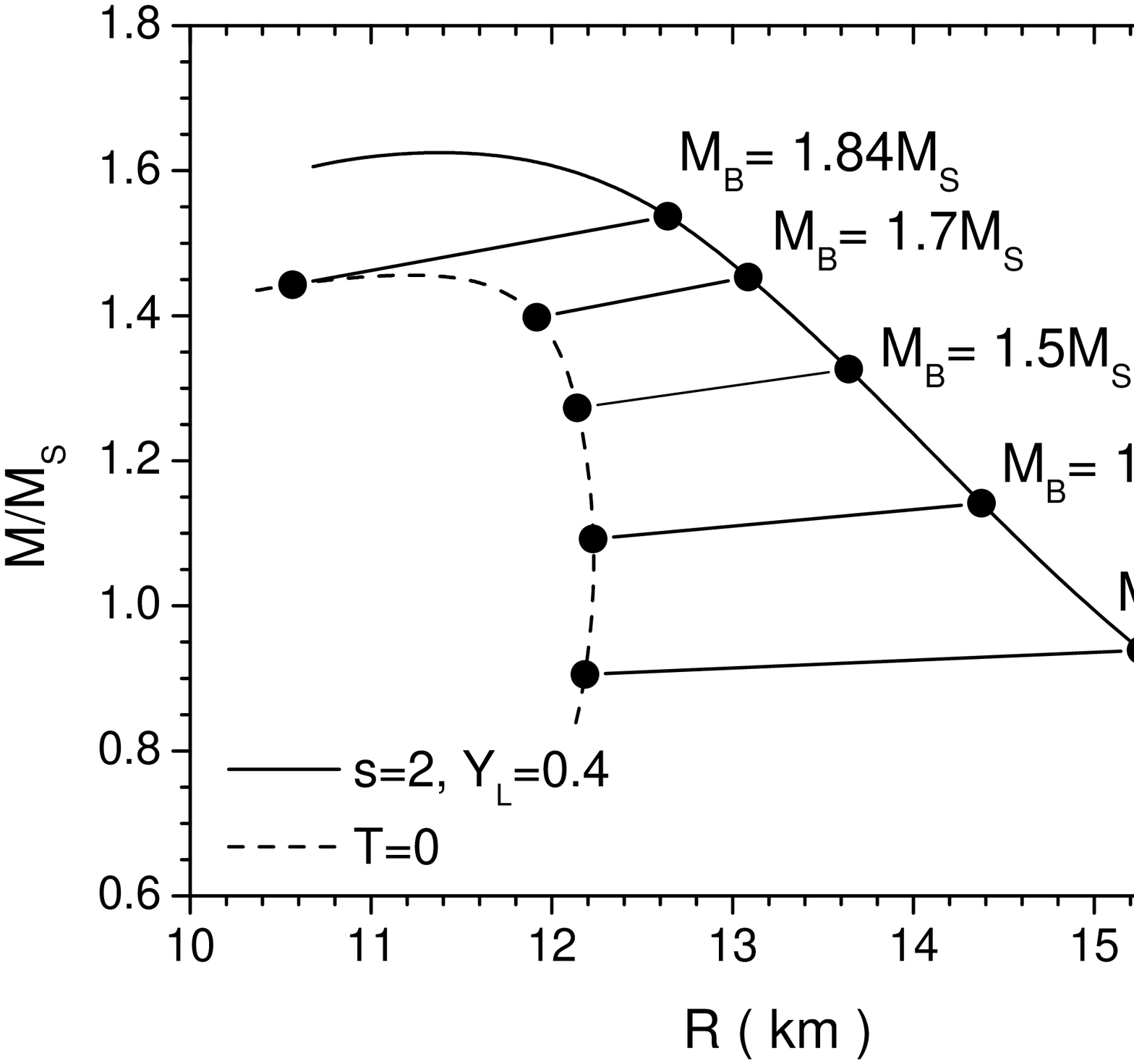}} ~~~~~~~~~~~~
\subfigure{\includegraphics[width=6cm]{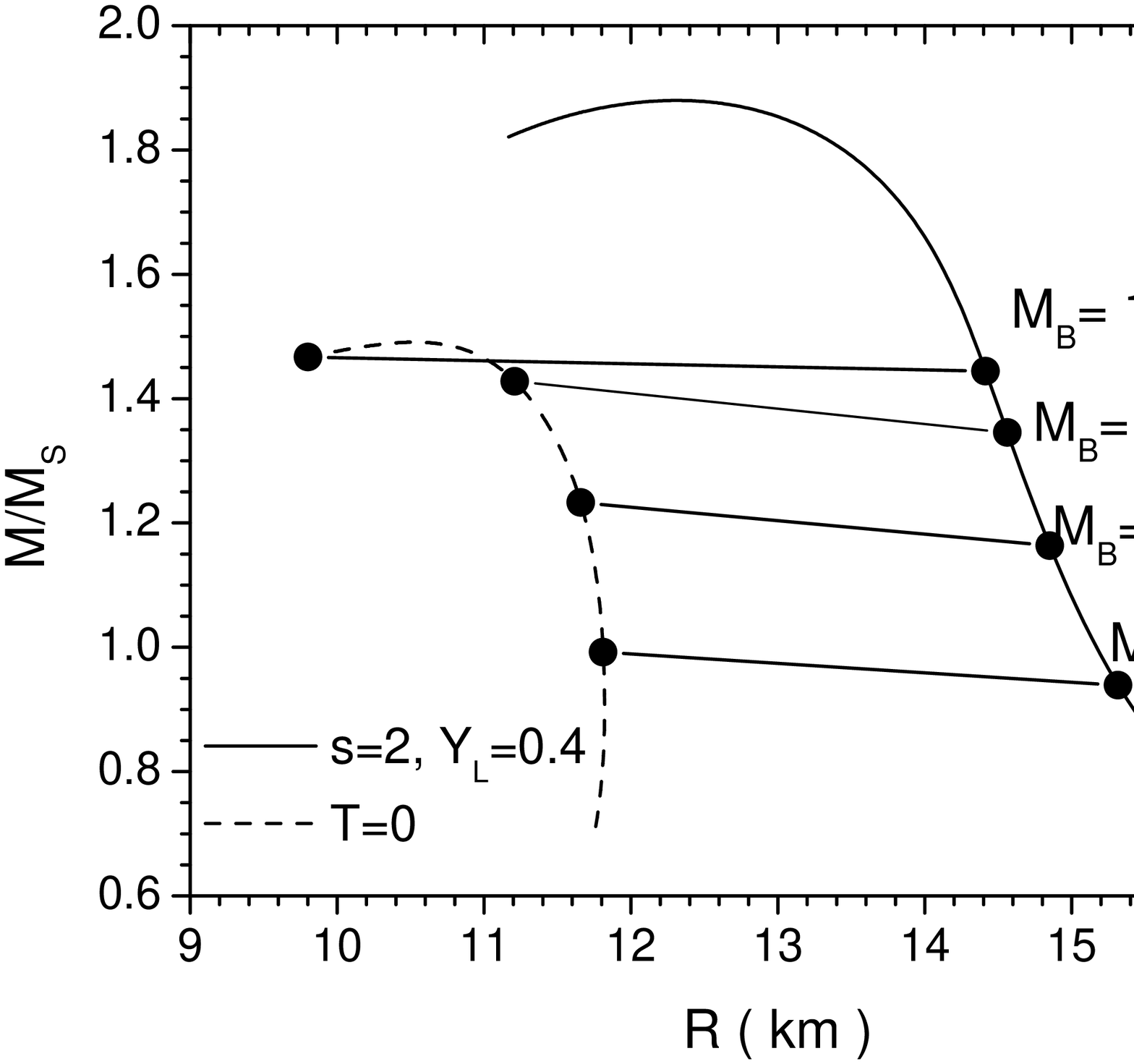}}
\end{center}
\caption{\it{Deleptonization of a protoneutron star. The left
panel presents results obtained for ZM1 parameter set and the
right panel for the SU3 model.}} \label{fig:ev}
\end{figure}

\begin{figure}[htbp]
\begin{center}
\subfigure{\includegraphics[width=6cm]{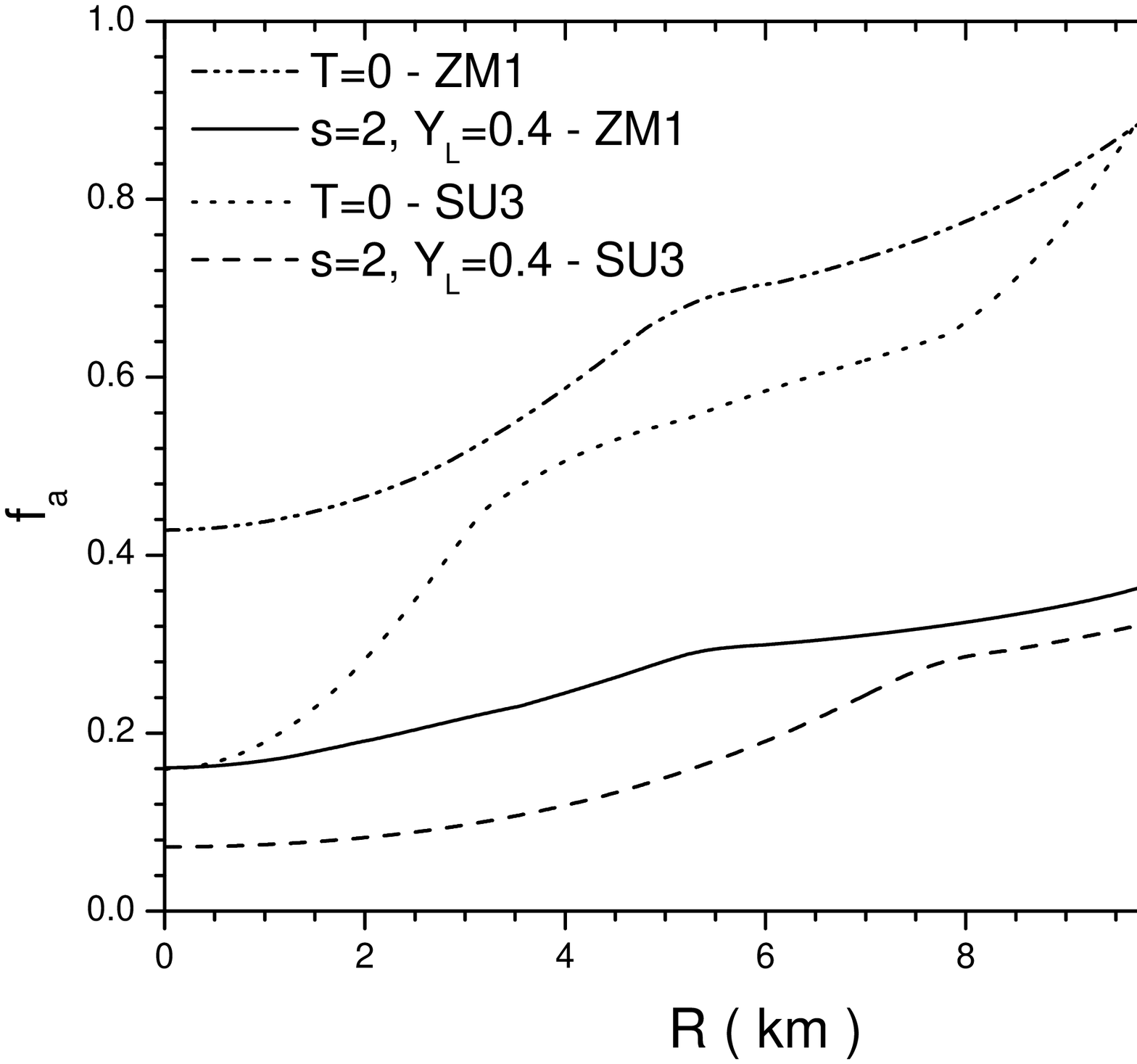}}~~~~~~~~~~~~
\subfigure{\includegraphics[width=6cm]{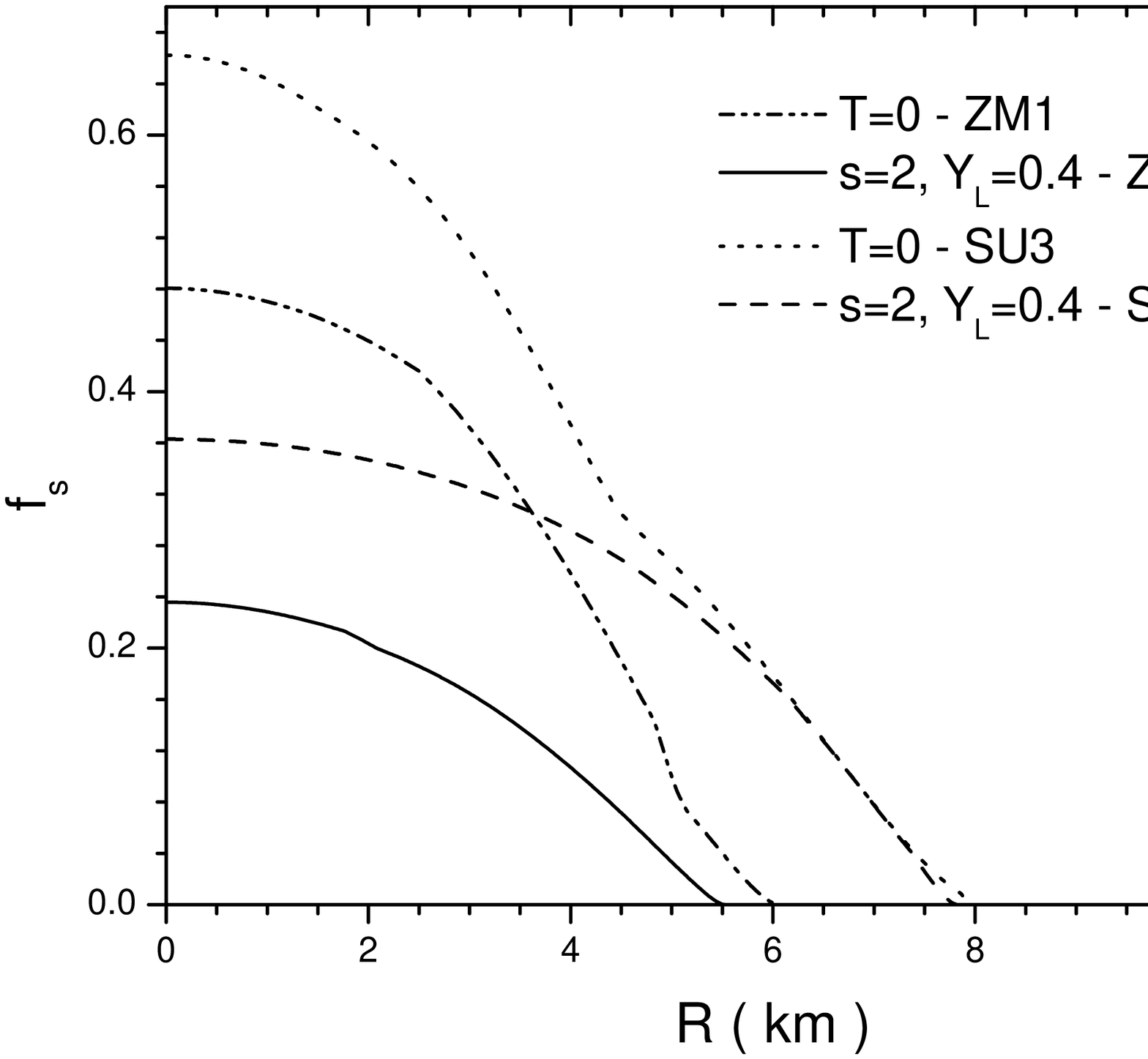}}
\end{center}
\caption{ {\it The asymmetry parameter $f_{a}$ (the left panel)
and the strangeness content parameter $f_{s}$  (the right panel)
as functions of the stellar radius $R$ .}}
 \label{fig:asym}
\end{figure}

\begin{figure}[htbp]
\begin{center}
\subfigure{\includegraphics[width=6cm]{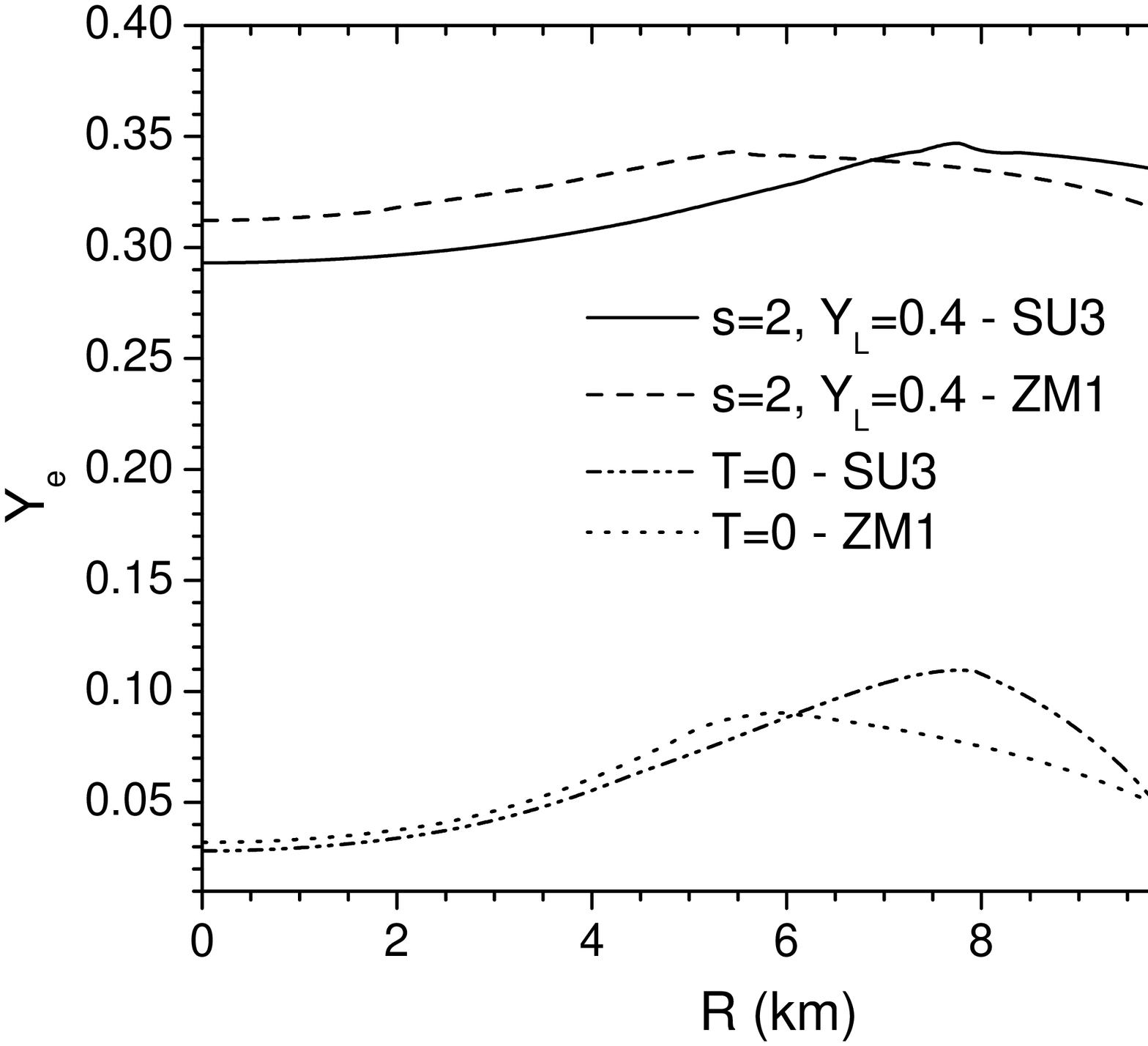}} ~~~~~~~~~~~~
\subfigure{\includegraphics[width=6cm]{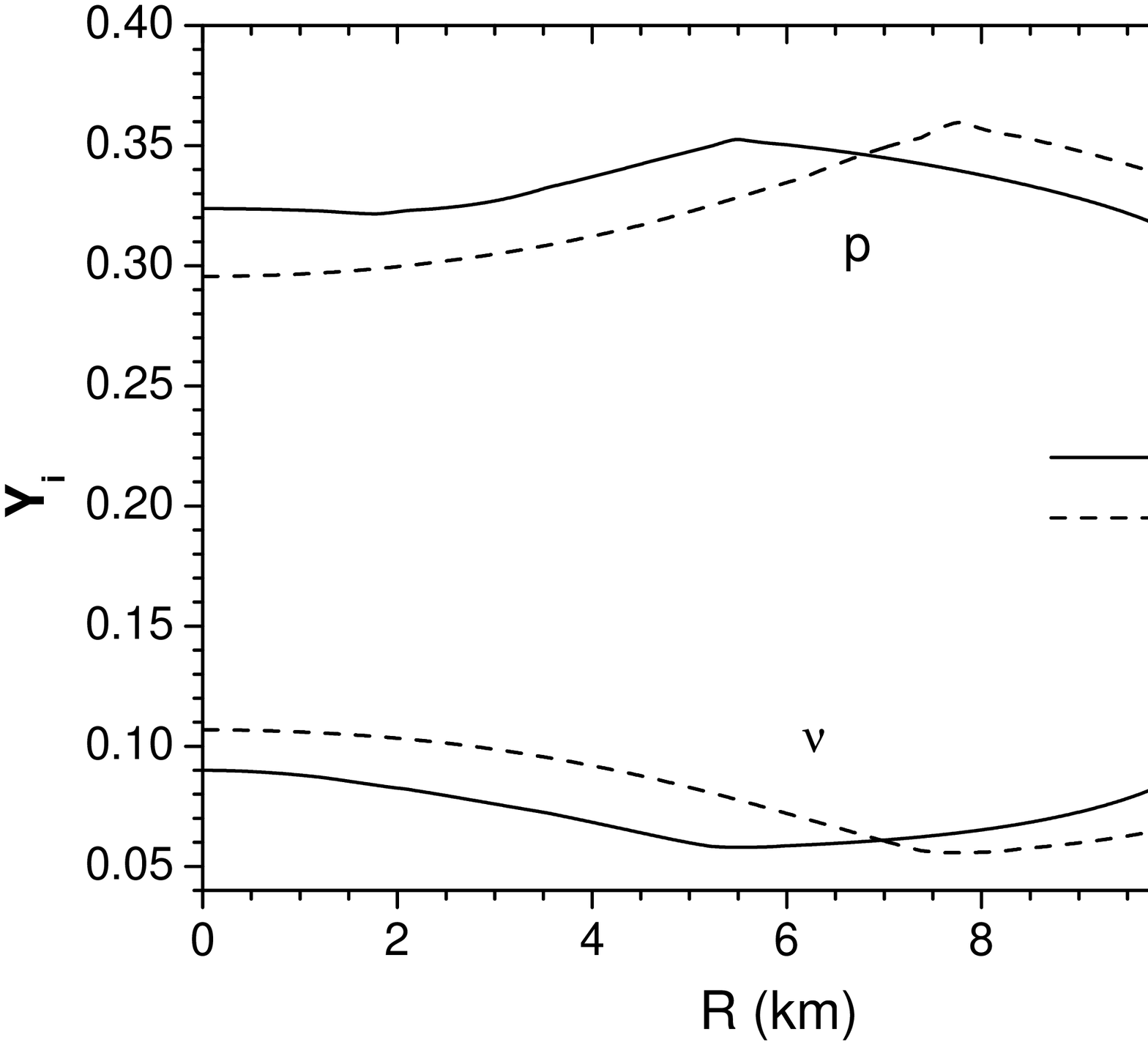}}
\end{center}
\caption{\it{Electron (left panel)and proton and neutrino (right
panel) fractions $Y_i$, as functions of baryon number density
$n_{B}$.}}
 \label{fig:lept}
 \end{figure}

\begin{figure}[htbp]
\begin{center}
\includegraphics[width=7cm]{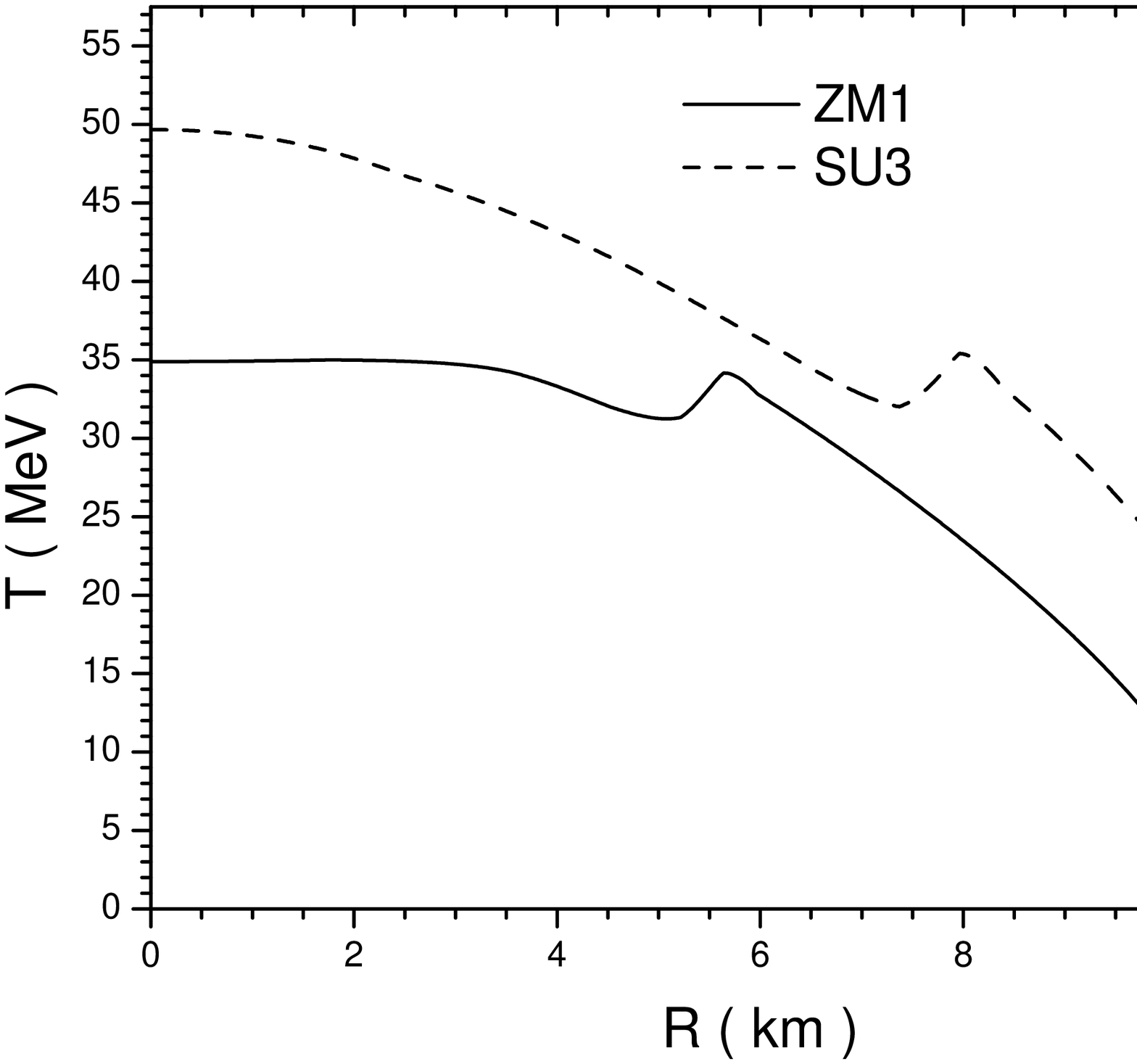}
\end{center}
\caption{\it {Stellar temperature as a function of the star
radius.}} \label{fig:tempstar}
\end{figure}

On obtaining  the form of the
equations of state of protoneutron star matter the mass-radius
relation and composition of the star can be specified. Two
distinctive parameter sets describing strangeness rich matter have
been used. The first one denoted as SU3 has been constructed on
the basis of chiral symmetry whereas the latter one (ZM1) has been
obtained with the use of the derivative coupling model. The
considered theories  have been extended by the inclusion of
$\delta$ meson and nonlinear vector meson interaction terms. The
inclusion of $\delta$ meson  seems to be indispensable for the
complete description of asymmetric neutron star matter.   The main
effect of the presence of $\delta$ meson becomes evident when
studying properties of  matter especially baryon mass splitting
and the form of the equation of state. Throughout the effective
baryon masses $\delta$ meson alters baryon chemical potentials
what realizes in characteristic modification of the appearance,
abundance and distribution of the individual flavors. The
assumptions underlying the calculations performed in this paper
are connected with the choice of the repulsive nucleon-hyperon
$\Sigma$ interaction.
\newline
The forms of the selected equations of state are shown in
Fig.\ref{fig:eos}.
In agreement with the generally accepted scheme of protoneutron
star evolution results for two different cases has been presented.
The first one corresponds to the era of neutrino trapping
($Y_L=0.4$ and $s=2$) and the second one fulfills the conditions
of cold, deleptonized matter ($Y_L = 0$ and $ s=0 $). This figure
presents also the influence of both entropy and neutrino trapping.
For these two cases the SU3 models lead to significantly stiffer
equations of state.
\newline
 In the four
successive panels of Fig.\ref{fig:partn} the analysis of the
relative concentrations of particles as functions of baryon number
density $n_B$ are presented.
These results indicate that the first strange baryon that emerges
is the $\Lambda$ it is followed by $\Xi^-$ and $\Xi^0$. For both
parameter sets  the sequence of appearance of hyperons is the same
however,in case of ZM1 parameter set, shifted towards higher
densities. Due to the repulsive potential of $\Sigma$ hyperons
their onset points are possible at very high densities which are
not relevant for neutron stars.  The appearance of $\Xi^-$
hyperons through the condition of charge neutrality affects the
electron and muon fractions and causes a drop in their contents.
Therefore the appearance of charged hyperons permits the lowering
of lepton contents and charge neutrality tends to be guaranteed
without lepton contribution. Larger effective baryon masses cause
the shift of given hyperon onset point especially for the charged
ones in the direction of higher densities.
\newline
Any realistic calculation of the properties of neutron stars is
based upon the general relativistic equation for hydrostatic
equilibrium (the Oppenheimer-Volkoff equation \cite{OVT}).
\newline
 On specifying  the equation of state
the solution of this equation can be found and this allows us to
determine global neutron stars parameters \cite{bedn}. The
structure of spherically symmetric neutron star is determined with
the use of this equation which is integrated starting from
$\rho=\rho_c$ at $r=0$ to the surface at $r=R$ where $\rho =0$. In
this way a radius $R$ and a mass $M=m(R)$ for a given central
density can be found . As a result, the mass-radius relation can
be drown. These relations for the chosen forms of the equations of
state are presented in Fig.\ref{fig:ev}. This particular sequence
of figures gives a detailed inside into what happens with lowering
number of neutrinos ($Y_{\nu}\rightarrow 0$) and decreasing value
of entropy. In the left panel of  Fig.\ref{fig:ev}  the
mass-radius relations for the  protoneutron star calculated with
the use of ZM1 parameter set is plotted. The right panel depicts
the mass-radius relations for SU3 model.  In general neutrino
trapping increases the value of the maximum mass. The SU3
parameter set gives higher value of the maximum mass then the ZM1
parameter set.
 Dots which are
connected by straight lines represent the evolution of
configurations characterized by the same baryon number. In both
figures  there are configurations which due to deleptonization go
to the unstable branch of the neutron star mass-radius relation.
The reduction in mass is much bigger in case of the SU3 parameter
set.
The stellar parameters obtained for the maximum mass configuration
for different models are presented in Table \ref{tab:maxmass}.
\begin{table}
\centering\begin{tabular}{|c|c|c|c|c|c|} \hline model & $M
(M_{S}$)& $R (km)$& $\rho_{c}$ ($10^{15} g/cm^{3}$)& $T_{c}$ (MeV) & $P_{c}$ $(MeV/fm^{3})$ \\
\hline SU3, $S=2$, $Y_{L}=0.4$ &
1.88 & 12.3 & 2.0 & 49.68 & 268.1\\ \hline SU3, $T=0$ & 1.49 & 10.52 & 2.43& 0.0 & 189.9\\
\hline ZM1, $S=2$, $Y_{L}=0.4$ & 1.62 & 11.41 & 2.44 & 34.88 & 300.4 \\
\hline ZM1, $T=0$ & 1.46 & 11.2 & 1.98& 0.0 & 166.0 \\ \hline
\end{tabular}
\caption{Properties of protoneutron maximum mass configurations
for SU3 and ZM1 parameter sets.} \label{tab:maxmass}
\end{table}
 The solutions of the structure equations also allows us  to carry out
an analysis of the onset point, abundance and distributions of the
individual baryon and lepton species   as functions of the star
radius. The comparison of results obtained for the two  cases
presented above (ZM1 and SU(3)) have been made on the basis of the
assumption that   the repulsive $\Sigma$ interaction shifts the
onset point of $\Sigma$ hyperons to very high densities.
Consequently,  they do not appear in the neutron star interior
calculated in these models. The maximum mass configurations has
been considered.
\begin{figure}[htbp]
\begin{center}
\subfigure{\includegraphics[width=6cm]{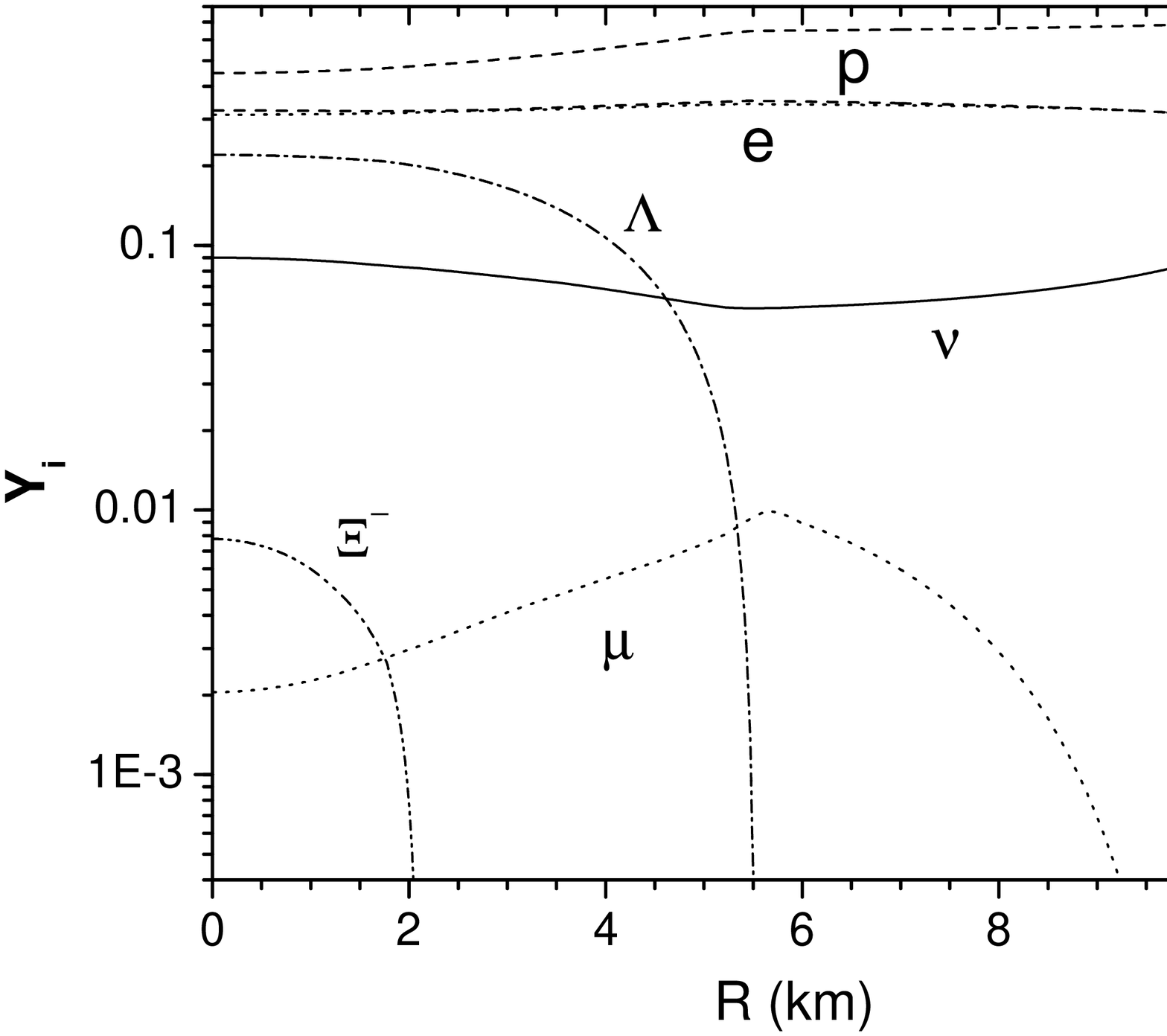}}~~~~~~~~~~~~
\subfigure{\includegraphics[width=6cm]{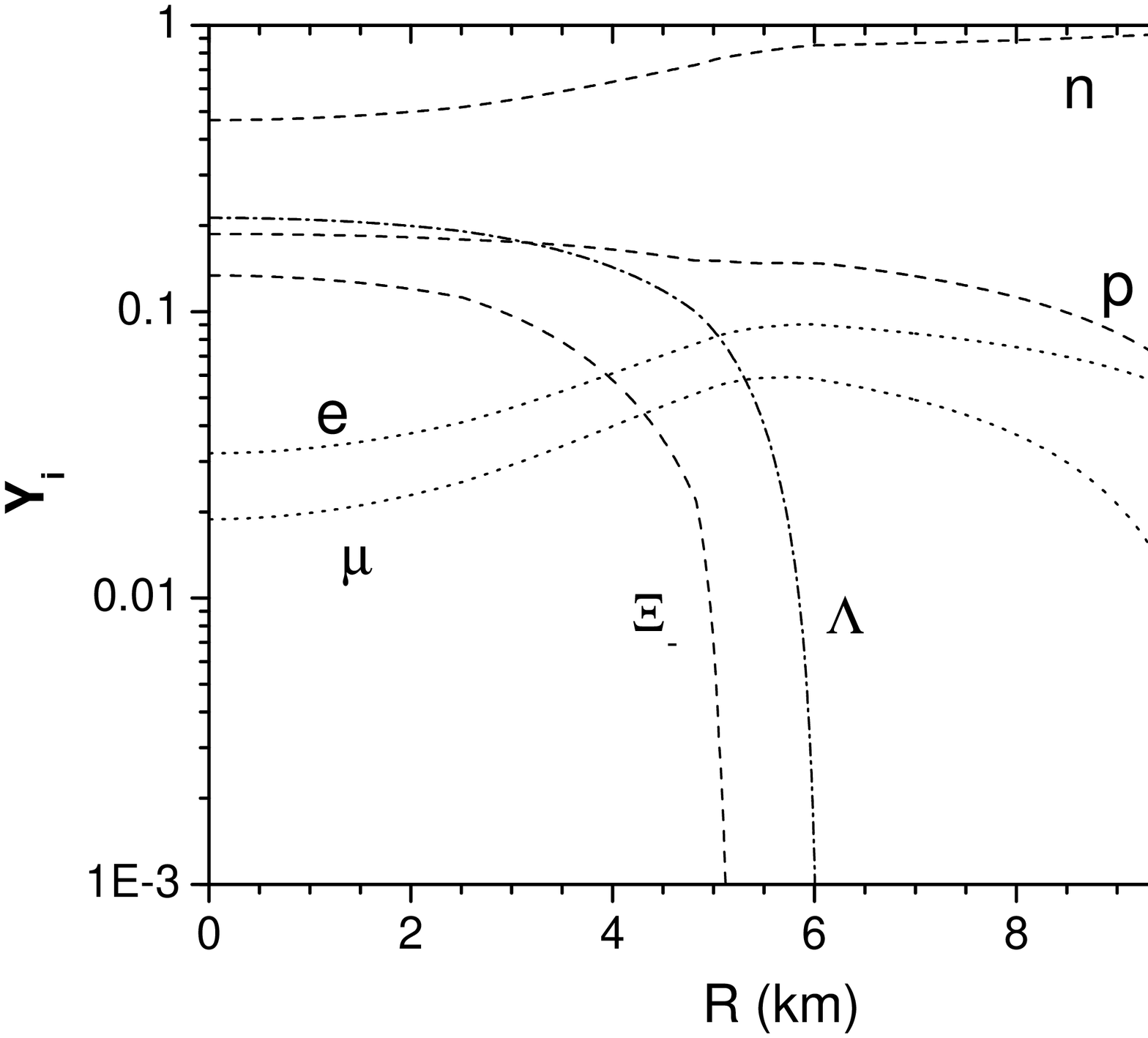}}
\subfigure{\includegraphics[width=6cm]{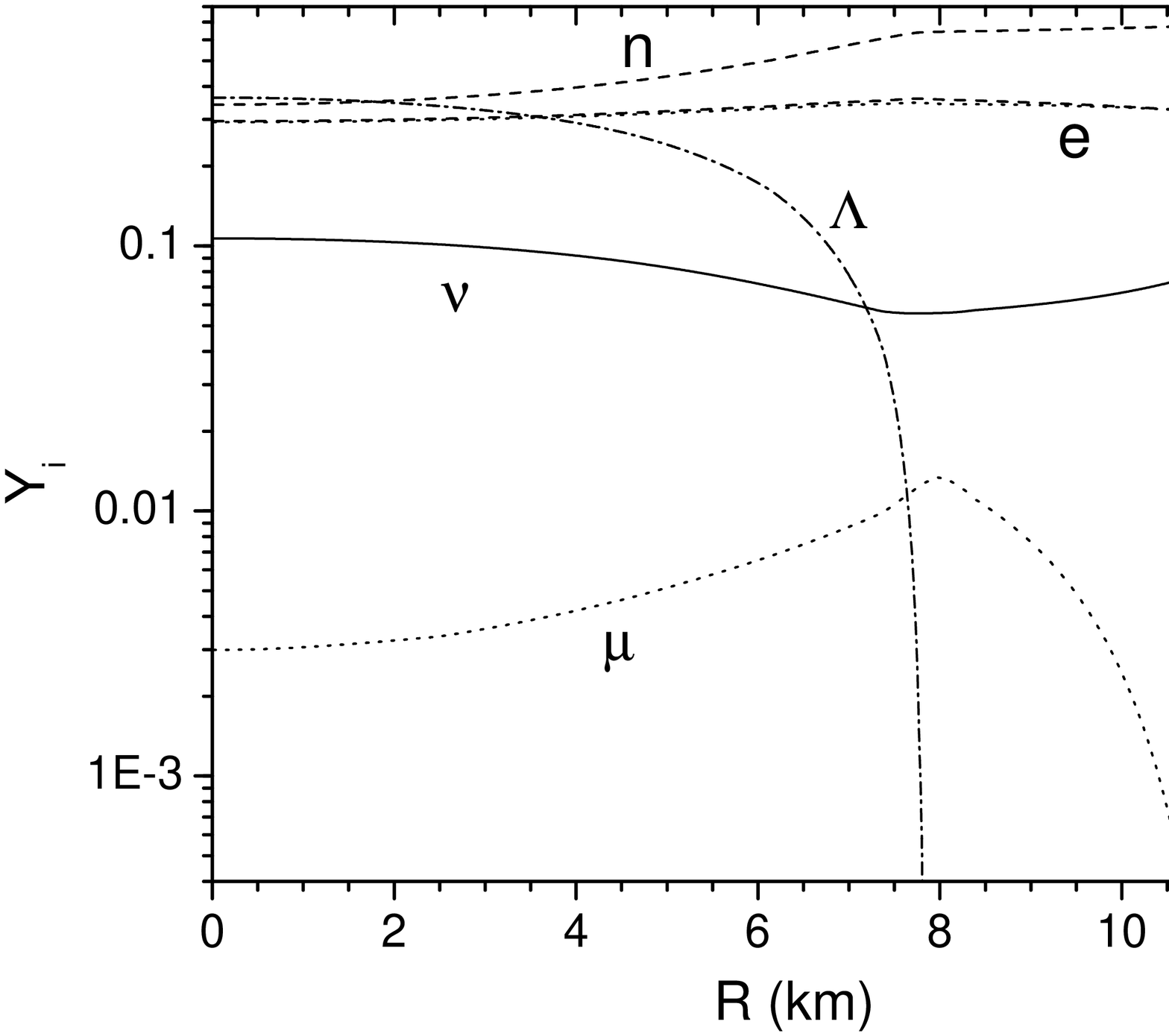}}~~~~~~~~~~~~
\subfigure{\includegraphics[width=6cm]{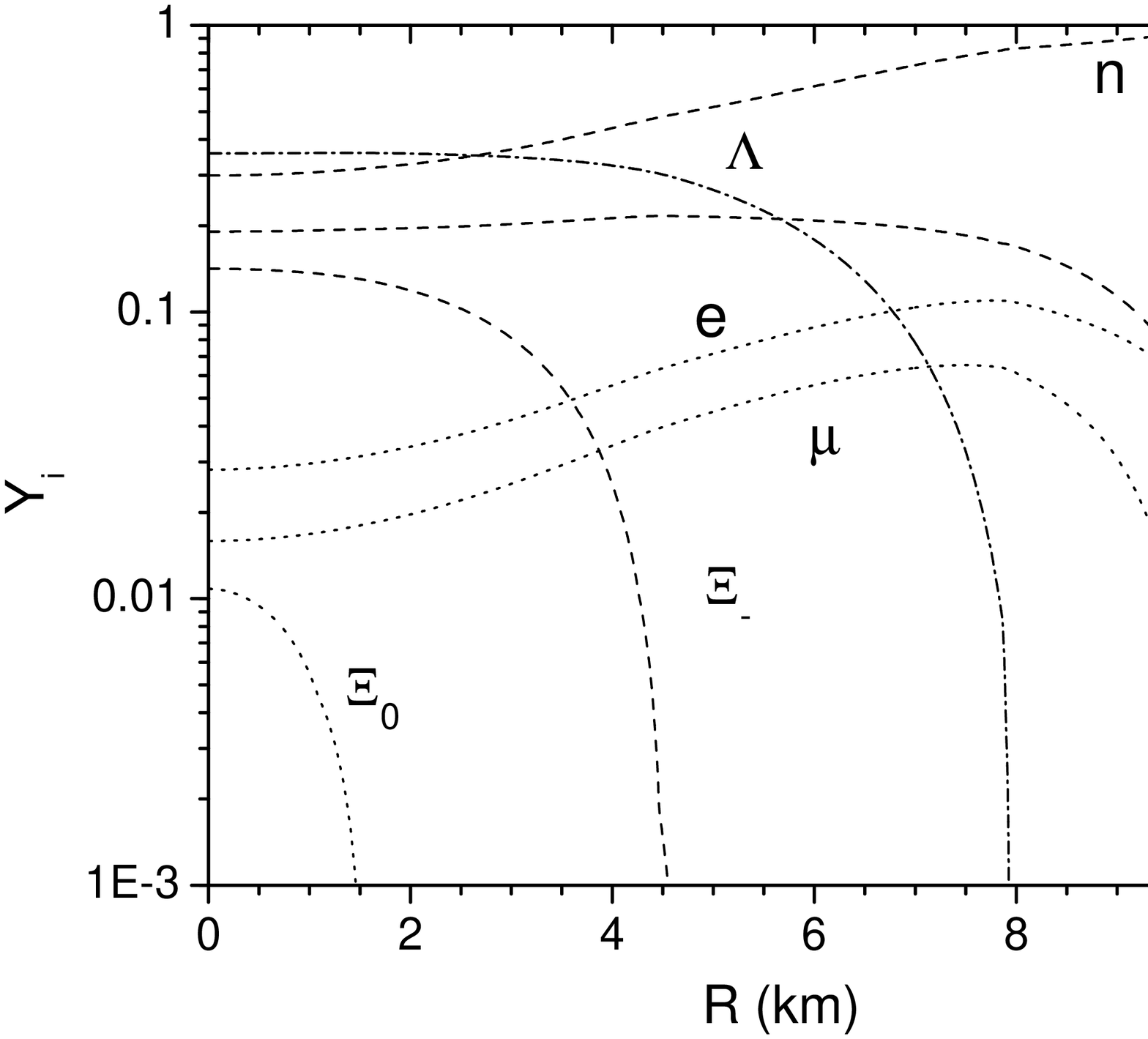}}
\end{center}
\caption{\it{The equilibrium composition  for the maximum mass
configuration as functions of the stellar radius $R$. Upper left
panel depicts the composition obtained on the basis of the ZM1
model for $s=2$ and $Y_l=0.4$. Upper right has been calculated for
$s=0$. Results obtained for SU3 model are presented in lower
panels $s=2$ and $Y_l=0.4$ (left) and  $s=0$ (right).}}
\label{fig:partr}
\end{figure}
In case of trapped neutrinos the ZM1 model leads to more compact
strangeness rich core. This very compact hyperon core which
emerges in the interior of the maximum mass configuration consists
of $\Xi^-$ and $\Lambda$ hyperons as shown in the upper left panel
of Fig.\ref{fig:partr}. The presence of negatively charged
hyperons reduces the content of negatively charged leptons. For
the SU3 model the  strangeness rich core is more extended (the
bottom left of Fig.\ref{fig:partr}) but in case of neutrino
trapped matter it contains only $\Lambda$ hyperon. The absence of
negatively charged hyperons leaves the electron and muon content
unchanged. After deleptonization the strangeness rich core in case
of the SU3 model remains more extended than in the case of ZM1
parameter set but it contains $\Xi^0$, $\Xi^-$ and $\Lambda$
hyperons.
\newline
The relative baryon composition in this model can be also analyzed
through the density dependence of the asymmetry parameter $f_a$
and  the strangeness contents $f_s$. Fig.\ref{fig:asym} presents
both parameters as functions of the star radius $R$.
As it was shown in Fig.\ref{fig:partr} trapped neutrinos have an
influence on the charged hyperon onset points. The appearance of
additional negatively charged particles has the consequence on the
proton content in the system and this in turn on the asymmetry
parameter $f_a$.
\newline
Fig. \ref{fig:lept} depict the relative fractions of protons,
electrons and muons for the ZM1 and SU3 parameter sets.
  Fig. \ref{fig:asym}
 compares the asymmetry parameter $f_a$ for ZM1 and the SU3
 parameter sets. It is noticeable that SU3 parameters give
 less asymmetric protoneutron  and neutron star but with the higher value of the strangeness content parameter $f_s$.
 \newline
 Fig. \ref{fig:tempstar} depicts the temperature profiles plotted
 for the presented above models.
In case of ZM1 protoneutron star model the temperature in interior
of maximum mass configuration takes higher value than in case of
SU3 model. The appearance of the strangeness-rich core on both
plots is marked by the local inversion of temperature.


\begin{thebibliography}
\expandafter\ifx\csname
natexlab\endcsname\relax\def\natexlab#1{#1}\fi
\expandafter\ifx\csname bibnamefont\endcsname\relax
  \def\bibnamefont#1{#1}\fi
\expandafter\ifx\csname bibfnamefont\endcsname\relax
  \def\bibfnamefont#1{#1}\fi
\expandafter\ifx\csname citenamefont\endcsname\relax
  \def\citenamefont#1{#1}\fi
\expandafter\ifx\csname url\endcsname\relax
  \def\url#1{\texttt{#1}}\fi
\expandafter\ifx\csname
urlprefix\endcsname\relax\def\urlprefix{URL }\fi
\providecommand{\bibinfo}[2]{#2}
\providecommand{\eprint}[2][]{\url{#2}}
\bibitem{aqu}Aguirre R., De Paoli A.L. 2002 \textit{Eur.Phys.J.} textbf{A13}, 501
\bibitem{bedn} Bednarek I., Manka R. 2001 \textit{Int. Journal
 Mod. Phys.} \textbf{D10}, 607
\bibitem{bedn2} Bednarek I., Keska M.,Manka R. 2003 \textit{Phys. Rev.} \textbf{C68},
035805
\bibitem{bodmer} Bodmer A.R., 1991, \textit{Nucl.Phys.}, \textbf{A526}, 703
\bibitem{bog77}J. Boguta and A.R. Bodmer, 1977, \textit{Nucl. Phys.}, \textbf{A292}
413, \textit{Int. J. Mod. Phys.}, \textbf{E6}, 515
\bibitem{bur}Burrows A., Lattimer J.M. 1986\textit{Astrophys. J.} \textbf{307} 178
\bibitem{GG:1969}Gasiorowicz S., Geffen D.A., 1969, \textit{Rev. Mod.
Phys.}, \textbf{41}, 531
\bibitem{glen}Glendenning N.K.,  \textit{Astrophys.J.}, \textbf{293}, 470 (1985); Also see in \textit{Compact
Stars} by N. K Glendenning Sringer-Verlag, New York (1997)
\bibitem{GM:1991}N.K. Glendenning and S.A. Moszkowski, 1991,
\textit{Phys. Rev. Lett.}, \textbf{67}, 2414
\bibitem{greiner}Greiner C., Schaffner-Biellich J.,
nucl-th/9801062
\bibitem{delta}Kubis S., Kutschera M., 1997, \textit{Phys. Lett.} \textbf{B399} 191
\bibitem{pra2}Lattimer J.M., Prakash M. 2001 \textit{Astrophys. J.}\textbf{550} 426
\bibitem{sua}Papazoglou P., Zschiesche D., Schramm S., Schaffner-Bielich J., Stöcker H., Greiner W., 1998, \textit{Phys.
Rev.},\textbf{C59},411,Papazoglou P.,Schramm S., Schaffner-Bielich
j., Stöcker H., Greiner W., 1998, \textit{Phys. Rev.},
\textbf{C57}, 2576, Hanauske M., Zschiesche D., Pal S., Schramm
S., Stöcker H., Greiner W.,2000,\textit{Ap.J.},\textbf{573},958
\bibitem{pra3}Pons J.A., Reddy S., Ellis P.J., Prakash M., Lattimer J.M. \textit{Phys. Rev.}\textbf{C 62} 035803
\bibitem{pra1}Prakash M., Bombaci I., Manju Prakash, Ellis P.J., Lattimer J.M., Knorren R. 1997 \textit{Phys. Rep.} \textbf{280} 1
\bibitem{rei}Reinhard P.-G., Rufa M. , Maruhn J., Greiner W.
and Friedrich J. 1986 \textit{Z. Phys. A - Atomic
Nuclei}\textbf{323} 13
\bibitem{dhsf}J. Schaffner and I.N. Mishustin,  1996, \textit{Phys. Rev.}, \textbf{C53}, 1416
\bibitem{ser}Serot B.D., Walecka J.D. 1986 \textit{Adv. Nucl. Phys.} \textbf{16} 1; 1997 \textit{Int. J. Mod. Phys.} \textbf{E6} 515
\bibitem{toki}Sugahara Y. and Toki H. 1994 \textit{Prog. Theo. Phys} \textbf{92}
803
\bibitem{OVT}R. C. Tolman, 1939, \textit{Phys. Rev.}, \textbf{55},
364; J.R. Oppenheimer and G.M. Volkoff, 1939, \textit{Phys. Rev.},
\textbf{55}, 374
\bibitem{NT}Törnqvist N.A., 1997, \textit{The linear U(3)$\times$
U(3) sigma model, the sigma (500) and the spontaneous breaking of
symmetries}, hep-ph/9711483
\bibitem{weber}Weber F. \textit{Pulsars as Astrophysical Laboratories for
Nuclear and Particle Physics}, 1999, IOP Publishing, Philadelphia
\bibitem{zima}Zimanyi J., Moszkowski S.A., 1990, \textit{Phys. Rev.} \textbf{C42} 1416
\bibitem{zschieshce} Zschiesche D., Mishira A., Schramm S.,
Stöcker H., Greiner W., nucl-th/0302073
\end{thebibliography}
\end{document}